\documentclass[onecolumn, aps,prx, showpacs, superscriptaddress, floatfix,nofootinbib]{revtex4-2}
\usepackage[T1]{fontenc}
\usepackage[utf8]{inputenc}
\usepackage{mathtools}
\usepackage{graphicx}
\usepackage{caption}
\usepackage{subcaption}
\usepackage{xcolor}
\usepackage{soul}
\usepackage{amsmath,latexsym}
\usepackage{array}
\usepackage{tikz,hyperref}
\usepackage{orcidlink}
\usepackage{float}

\newcommand{\msun}{M$_\odot$}
\newcommand{\ft}[0]{\footnotesize}
\newcommand{\physrep}{Phys.~Rep.}
\newcommand{\mnras}{Mon.~Not.~Roy.~Astron.~Soc.}
\newcommand{\apjl}{Astrophys.~J.~Lett.}
\newcommand{\pasp}{PASP}
\newcommand{\aapr}{Astron.~Astrophys.~Rev.}
\newcommand{\araa}{Annu.~Rev.~Astron.~Astrophys.}
\newcommand{\apjs}{Astrophys.~J.~Suppl.~Ser.}

\begin{document}

\title{Hyperons and $\Delta$'s in rotating protoneutron stars: Global properties}

\author{Franciele M. da Silva \orcidlink{0000-0003-2568-2901}} 
\email{franciele.m.s@ufsc.br}

\affiliation{Departamento de F\'isica, CFM - Universidade Federal de Santa Catarina; \\ Caixa Postal 476, CEP 88.040-900, Florian\'opolis, SC, Brazil.}

\author{Adamu Issifu \orcidlink{0000-0002-2843-835X}} 
\email{ai@academico.ufpb.br}
\affiliation{Departamento de F\'isica e Laborat\'orio de Computa\c c\~ao Cient\'ifica Avan\c cada e Modelamento (Lab-CCAM), Instituto Tecnol\'ogico de Aeron\'autica, DCTA, 12228-900, S\~ao Jos\'e dos Campos, SP, Brazil} 

\author{Luis C. N. Santos \orcidlink{0000-0002-6129-1820}}
\email{luis.santos@ufsc.br}

\affiliation{Departamento de F\'isica, CFM - Universidade Federal de Santa Catarina; \\ C.P. 476, CEP 88.040-900, Florian\'opolis, SC, Brazil.}

\author{Tobias Frederico \orcidlink{0000-0002-5497-5490}} 
\email{tobias@ita.br}

\affiliation{Departamento de F\'isica e Laborat\'orio de Computa\c c\~ao Cient\'ifica Avan\c cada e Modelamento (Lab-CCAM), Instituto Tecnol\'ogico de Aeron\'autica, DCTA, 12228-900, S\~ao Jos\'e dos Campos, SP, Brazil}

\author{D\'ebora P. Menezes \orcidlink{0000-0003-0730-6689}}
\email{debora.p.m@ufsc.br}

\affiliation{Departamento de F\'isica, CFM - Universidade Federal de Santa Catarina; \\ C.P. 476, CEP 88.040-900, Florian\'opolis, SC, Brazil.}

\begin{abstract}
    Rotation plays an important role in the evolution of most types of stars, in particular, it can have a strong influence on the evolution of a newly born proto-neutron star. In this study, we investigate the effects of rotation on four snapshots of the evolution of proto-neutron stars with hyperons and $\Delta$-resonances in their cores, from birth as neutrino-rich objects to maturity as cold, catalyzed neutron stars. We focus on the effects of uniform rotation on the macroscopic structure of the star at three rotational frequencies -- 346.53 Hz, 716 Hz, and the Kepler frequency. Our investigation indicates that the impact of rotation at frequencies of $346.53$~Hz and $716$~Hz causes minor changes in the maximum gravitational mass but leads to significant changes in the stellar radius, particularly for stars with masses smaller than $2$~\msun. However, we observe drastic changes in the star's mass and radius when considering the Kepler frequency. In addition, we investigate other relevant characteristics of the rotating proto-neutron stars as they evolve such as the moment of inertia, compactness, central temperature, and Kerr parameter. Our results suggest that the inclusion of new degrees of freedom in the stellar core lead the star to be more sensitive to rotational dynamics, owing to an increase in compactness, a decrease in the central temperature, and a decrease in the moment of inertia.
\end{abstract}

\maketitle

\section{Introduction}

When massive stars exhaust their nuclear fuel and eventually reach the end of their life, their cores collapse, often resulting in a supernova explosion and the formation of either a neutron star (NS) or a black hole (BH). Supposing the supernova does not result in the formation of a BH, the dense core of the progenitor massive star initially forms a proto-neutron star (PNS). The newly born PNS is typically hot, lepton-rich, and rapidly rotating. After the ejection of the stellar envelope during the supernova explosion, deleptonization occurs, releasing the trapped neutrinos. The process of deleptonization continues through neutrino diffusion, heating the star while decreasing the net lepton fraction. Afterward, the star enters a neutrino-poor phase and {slowly cools down} until it forms a cold-catalyzed NS several years later~\cite{Pons:1998mm, Prakash:1996xs,burrows1986}. This way, the PNSs are believed to evolve in four main stages, as follows: lepton-rich stage, neutrino-poor stage, cooling phase, and the formation of cold-catalyzed NSs~\cite{Prakash:1996xs, burrows1986}. These processes are prominently governed by the properties of dense matter under extreme temperature and density conditions, along with neutrino reaction rates and diffusion timescales. A detailed study of the microphysics properties of PNSs can be found in Ref.~\cite{Pons:2001ar, Pons:2000xf, keil1995}.

Rotation is another important ingredient in the evolutionary stages of a PNS. Observational evidence shows that massive stars that end their lives through supernova explosions are usually rapid rotators \cite{vanbelle2012A&ARv,fukuda1982PASP}. Moreover, there is strong evidence that long-duration gamma-ray bursts are associated with the formation of magnetars rotating at frequencies near the Kepler limit, or with a precursor stage to the formation of a rapidly rotating BH~\cite{langer2012ARA&A, woosley2006ARA&A}. Thus, the shrinking core that results from the explosion of a massive star is thought to be rapidly rotating. The numerical simulations of core collapse and explosion \cite{nakamura2014ApJ, mosta2014ApJ, summa2018ApJ, thompson2005ApJ} and later evolutionary stages of massive stars \cite{heger2005ApJ, hirschi2004A&A} provide estimates for the minimum period of NSs at birth to be $\sim 1-15$~ ms; however, these models still present many uncertainties. However, during the evolution of the PNSs, a great amount of angular momentum can be lost due to neutrino emission~\cite{janka2004,martinon2014PhRvD}, gravitational radiation~\cite{andersson2003CQGra}, magnetic winds~\cite{heger2005ApJ,thompson2004ApJ} and other possible mechanisms~\cite{ott2006ApJS}. In this way, modeling the evolution of PNSs can be a very complex task, since it may involve the incorporation of neutrino transport in the equation of state (EoS)~\cite{marek2009ApJ,takiwaki2014ApJ,summa2018ApJ} and general relativistic hydrodynamics that account for fluid motion in rotating relativistic stars~\cite{thompson2005ApJ,summa2018ApJ} as well as the effects of strong magnetic fields~\cite{franzon2016,heger2005ApJ,mosta2014ApJ}. Therefore, the microphysics inputs need to be carefully treated for consistency. As discussed in~\cite{Pons:1998mm}, the careful construction of the EoS for dense nuclear matter, neutrino opacity, and the interactions between neutrinos and matter are important factors for achieving reliable results. 

Some efforts have been made to simplify the modeling of the PNSs in the literature~\cite{Dexheimer:2008ax, Issifu:2023qyi, Malfatti:2019tpg, Raduta:2020fdn, Pons:1998mm, Prakash:1996xs, Ghosh:2023tbn}, to obtain an insight into the structure of the PNSs, without loss of generality, using quasistationary approximation, imposing some {\it ad hoc} thermodynamic constraints (such as fixed temperature, fixed entropy per baryon and/or fixed lepton fraction) we can study the PNSs evolution as a sequence of equilibrium
configurations. The quasistationary assumption is justified because, during the early life of a PNS, it quickly attains equilibrium among gravitational, pressure, and rotational forces on a timescale much shorter than its overall evolutionary timescale. Other slow processes govern its long-term evolution, such as neutrino diffusion, cooling via thermal radiation emission, and rotational evolution (like angular momentum redistribution, viscosity, and magnetic fields)~\cite{Villain:2003ey, Prakash:1996xs, burrows1986}. While the quasistationary assumption is useful and justified for simplifying PNS models by focusing on the long-term timescale, it also has known limitations that can affect the accuracy of simulations and theoretical predictions. It does not account for essential physical phenomena that occur on shorter timescales or under extreme conditions. Addressing these issues requires advanced computational efforts and sophisticated modeling, leaving room for further development in non-equilibrium modeling. In this work, we adopt the rigid rotation approximation, assuming that the star rotates as a solid body with a constant angular velocity at each stage of its evolution. This implies the absence of differential rotation, as all parts of the star share the same angular velocity.

Advancements in observational astrophysics and increased data availability have imposed stringent constraints on the dense matter EoS through measurements of the maximum possible gravitational masses and radii of NSs. The recent data from NICER X-ray observatory has led to simultaneous measurement of both the mass and radius of the pulsars: PSR J0030$+$0451~\cite{riley2019, Miller:2019cac}, PSR J0740$+$6620~\cite{riley2021,miller2021}, PSR J0437$-$4715~\cite{choudhury2024ApJ} and PSR J1231$-$1411~\cite{salmi2024ApJ} with good precision. Additionally, data from the binary NS merger event GW170817, which led to the observation of the gravitational wave signal~\cite{LIGOScientific:2017vwq} together with the electromagnetic counterpart~\cite{2017ApJ...848L..12A,2017ApJ...848L..13A}, has provided valuable insights into NS properties~\cite{gw172018}. These constraints serve as a benchmark for modeling PNSs. Even though the dynamics of hot EoSs differ significantly from those of cold EoSs, a realistic and consistently modeled PNS EoS must demonstrate satisfaction with these constraints when the star is cold and catalyzed. The EoS intended for this work has been proven to satisfy the constraints mentioned above when the star is cold~\cite{Issifu:2023qyi, Issifu_2025}.

In this work, we employ an EoS model to study the Kelvin-Helmholtz phase, such that we use the assumption of a quasistationary evolution to study the PNSs from birth to maturity imposing thermodynamic conditions such as fixed entropy per baryon and fixed lepton fraction. The EoSs have been tested in studying PNSs with exotic baryons in~\cite{Issifu:2023qyi}, PNSs with quark cores in~\cite{Issifu_2025} and the hybrid NSs in~\cite{Issifu:2023ovi}. In the present work, we assume the PNS is a rigidly rotating compact object~\cite{Goussard:1996dp} and investigate its observable properties at different frequencies. Rigid rotation allows NSs to exist in a supermassive state~\cite{cook1992ApJ}, where their gravitational mass exceeds the non-rotating maximum mass limit due to rotational support. This phenomenon provides valuable insights into achieving higher maximum stellar masses, enhanced stability, and observable properties such as oblateness and rapid spin rates. These characteristics are crucial for understanding astrophysical phenomena, including pulsars, gravitational wave sources, and merger events.
On the other hand, the long-term effects of rigid rotation are influenced by the interplay between energy dissipation and angular momentum loss dynamics. In rigid rotation, the angular velocity remains constant throughout the entire star, whereas in non-rigid rotation~\cite{Bozzola:2017qbu, Galeazzi:2011nn}, it varies along the radius. This distinction simplifies calculations for the case of rigid rotation and offers valuable insights into the star's structural variations caused by rotation throughout its evolutionary path~\cite{Paschalidis:2016vmz, Villain:2003ey}.

The paper is organized as follows: In Sec.~\ref{mp}, we discuss the microphysics that governs the calculations of the EoS, where we introduce the DDME2 parameterization and its extension to heavy baryons. The structure of the spacetime metric associated with relativistic, rigidly rotating stars with isentropic matter is considered in Sec. \ref{stellar}.  In Sec. \ref{results}, we present the analyses of the impact of rotation on the macroscopic structure of the PNSs. Finally, in Sec. \ref{remarks}, we present our final remarks and conclusions.

\section{Microphysics}\label{mp}

The interaction between hadronic particles is governed by quantum hadrodynamics, where the strong nuclear force between the hadrons is modeled through the exchange of massive mesons. We consider the mesons represented by the fields \(i=\sigma, \omega, \rho, \phi\), where \(\sigma\) is a scalar meson, \(\omega\) and \(\phi\) are vector-isoscalar mesons (with \(\phi\) carrying hidden strangeness), and \(\rho\), a vector-isovector meson. They mediate the interaction. The associated Lagrangian density is given by:
\begin{equation}
     \mathcal{L}= \mathcal{L}_{\rm H}+ \mathcal{L}_\Delta + \mathcal{L}_{\rm m}+ \mathcal{L}_{\rm L} ,
\end{equation}
where $\mathcal{L}_{\rm H}$ represents the baryon octet, $ \mathcal{L}_\Delta$ represents the $\Delta$-resonances, $\mathcal{L}_{\rm m}$ represents the {mass terms of the} mediating mesons and $\mathcal{L}_{\rm L}$ represents the free leptons. The individual Lagrangian densities can be expressed explicitly as:              
\begin{align}
 \mathcal{L}_{\rm H}{}&=  \sum_{b}  \bar \psi_b \Big[  i \gamma^\mu\partial_\mu - \gamma^0  \big(g_{\omega b} \omega_0  +  g_{\phi b} \phi_0+ g_{\rho b} I_{3b} \rho_{03}  \big)
 - \Big( m_b- g_{\sigma b} \sigma_0 \Big)  \Big] \psi_b, \label{h}\\
   \mathcal{L}_{\Delta}{}&= \sum_{d}\Bar{\psi}_{d}\Big[i \gamma^\mu \partial_\mu- \gamma^0\left(g_{\omega d}\omega_0 + g_{\rho d} I_{3d} \rho_{03} \right) -\left(m_d-g_{\sigma d}\sigma_0 \right)\Big]\psi_{d},\label{d}\\
\mathcal{L}_{\rm m}&= - \frac{1}{2} m_\sigma^2 \sigma_0^2  +\frac{1}{2} m_\omega^2 \omega_0^2  +\frac{1}{2} m_\phi^2 \phi_0^2 +\frac{1}{2} m_\rho^2 \rho_{03}^2,\label{m}\\
 \mathcal{L}_{\rm L}& = \sum_L\Bar{\psi}_L\left(i\gamma^\mu\partial_\mu-m_L\right)\psi_L\label{l},
\end{align}
in natural units, where $c=\hbar=1$. Here, \(\psi_b\) is the baryonic Dirac-type field, with the subscript $b$ indicating summation over all baryon octets (nucleons and hyperons), $g_{ib}$ denotes the baryon-meson couplings, and $I_{3b} = \pm 1/2$ represents the isospin projection in Eq.~(\ref{h}). The $\Delta$-resonances are described by the Rarita-Schwinger-type Lagrangian density, represented by $\psi_d,$ where $d$ indexes all the $\Delta$-resonances, and $I_{3d} = \pm 3/2 $ denotes the isospin projections in Eq.~(\ref{d}). This formulation for the $\Delta$-resonances is used because their vector-valued spinor contains additional components compared to the four components of the spin-\(\frac{1}{2}\) Dirac-type spinors \cite{dePaoli:2012eq}. Additionally, Eq.~(\ref{m}) represents the Lagrangian density of the mediating massive mesons, with $m_i$ representing the meson masses presented on Tab.~\ref{T} and the subscript `0' representing the mean-field version of the meson fields. Finally, Eq.~(\ref{l}), is the Lagrangian density of the free leptons described by the Dirac free Lagrangian density. The summation index $L$ runs over all the lepton flavors in the stellar matter at each stage of the stellar evolution. For cold and catalyzed stellar matter, \(L\) includes electrons (\(e\)) and muons (\(\mu\)), while \(\tau\)-leptons are considered too heavy to be present. In the case of fixed entropy per baryon and fixed lepton fraction, $e$'s and their corresponding lepton neutrinos ($\nu_e$) are considered since 
{simulations indicate that} the presence of muons in supernova remnant matter becomes relevant only after the matter becomes neutrino transparent following supernova physics \cite{Pons:1998mm, Prakash:1996xs}. The presence of hyperons and \(\Delta\)-resonances is considered because neutrons, being fermions governed by the Pauli exclusion principle, experience increasing degeneracy pressure at higher densities. As a result, it becomes energetically favorable for some neutrons to convert into heavier baryons, such as hyperons and \(\Delta\)-resonances, to alleviate this pressure \cite{Bombaci:2016xzl, Menezes:2021jmw}.

The EoS is calculated using the density-dependent parameterization, commonly referred to in the literature as the DDME2 parameterization \cite{Roca-Maza:2011alv}. The meson-nucleon couplings are adjusted through the expressions:
\begin{align}
    g_{i b} (n_B) &= g_{ib} (n_0)a_i  \frac{1+b_i (\eta + d_i)^2}{1 +c_i (\eta + d_i)^2}, \label{cp}\\
    g_{\rho b} (n_B) &= g_{\rho b} (n_0) \exp\left[ - a_\rho \big( \eta -1 \big) \right],
\end{align}
where $n_B$ is the total baryon density and $\eta = n_B/n_0$. Under this framework the free model parameters: $a_i,\, b_i,\, c_i$ and $d_i$ are fitted to experimental bulk nuclear properties with saturation density $n_0=0.152~\rm fm^{-3}$ with the following properties: $E_B = -16.14$~MeV~(binding energy), $K_0 = 251.9$~MeV~(incompressibility), $J = 32.3$~MeV~(symmetry energy), and $L_0 =51.3$~MeV~(symmetry energy slope). These listed properties agree with recent constraints on the characteristics of symmetric nuclear matter, as reported in Refs.~\cite{Reed:2021nqk, Lattimer:2023rpe, Dutra:2014qga}. The determined model parameters are presented in Tab.~\ref{T}.

The fitting of the meson-nucleon couplings discussed above was through fitting to pure nucleonic matter. An extension to include the hyperons and $\Delta$-resonances couplings is done through several approaches reported in the literature \cite{Glendenning:1991es, Lopes:2020rqn, Pais:1966eox, Weissenborn:2011kb}. However, in the present work, we use the meson-baryon coupling reported in \cite{Lopes:2022vjx}, where the authors determine the couplings using SU(3) and SU(6) symmetry arguments and potential depths of each particle. One of the advantages of using these couplings is that they enable us to obtain stellar masses within the $2$~\msun~threshold in the presence of hyperons and $\Delta$-resonances. The adopted coupling parameters are presented in Tab.~\ref{T1}. 

{In Ref.~\cite{sedrakian2022}, the authors also considered meson–hyperon couplings based on SU(6) flavor symmetry arguments, with magnitudes comparable to those reported in \cite{Lopes:2022vjx}. However, in \cite{sedrakian2022}, they adopted a universal coupling scheme for all meson–$\Delta$ resonance interactions by setting the coupling ratios to unity. In contrast, in \cite{Lopes:2022vjx}, the meson–$\Delta$ couplings were determined analogously to the meson–hyperon couplings, using a combination of SU(3) and SU(6) symmetry arguments. As discussed in Refs.~\cite{Li:2019tjx, Schurhoff:2010ph}, the emergence of the $\Delta^-$ resonance in dense matter alters the chemical equilibrium conditions among the particles involved in the Urca process, $n \rightarrow p + e^- + \bar{\nu}_e$. This rearrangement of chemical potentials leads to an enhanced proton fraction compared to matter composed solely of $npe\mu$. The presence of the $\Delta^-$ can also influence neutrino diffusion and, consequently, the thermal evolution of the star by introducing additional weak interaction channels, such as $\Delta^- \rightarrow n + e^- + \bar{\nu}_e$, or, in the presence of hyperons, $\Delta^- \rightarrow \Lambda + e^- + \bar{\nu}_e$ \cite{Prakash:1992zng}.}

The details and relevant EoS under this formalism are presented in \cite{Issifu:2023qyi}, where the EoS was originally used to study stellar evolution from supernova remnants. Further developments of the EoS can be found in \cite{Issifu_2025, Issifu:2024htq}, where these EoSs were applied to investigate other astrophysical phenomena, such as hybrid PNSs and the effects of dark matter on stellar evolution. Additionally, similar derivations of these EoSs are detailed in Ref.~\cite{Malfatti:2019tpg, Raduta:2020fdn, sedrakian2022}, and we do not intend to repeat them here. Instead, we highlight some relevant relations that connect the entropy density ($s$), the EoS (energy density ($\varepsilon$) versus pressure ($p$)), and the temperature ($T$) fluctuations within the star in the neutrino-trapped and neutrino-transparent regimes of stellar evolution. The thermodynamic relation connecting these variables is given by the expression for the free energy density: $\mathcal{F}_B = \varepsilon_t - Ts$, where $\varepsilon_t$ represents the total energy density of the system. Therefore, the explicit relation connecting the $s, T, p_t$ ($p_t$ is the total pressure) and $\varepsilon_t$ becomes:
\begin{align}
     sT= \varepsilon_t + p_t- \sum_{b} \mu_{b} n_{b} -\sum_{d} \mu_{d} n_{d} -\sum_{L} \mu_{L} n_{L}, 
\end{align}
where $\varepsilon_t$ and $p_t$ include the sum of the pressures and the energy densities of all the particles in the system, \(\mu\) is the chemical potential, and \(n\) is the number density, the subscripts denote the particle type, and the sum extends over all particles of that type. Since the NSs are physically observable objects, the stellar matter is in $\beta$-equilibrium and charge neutrality. Applying these conditions, the above expression can be simplified further:
\begin{equation}
    sT=p_t+\varepsilon_t -n_B\mu_B,
\end{equation}
for neutrino-transparent matter, where $\mu_B$ is the total baryon chemical potential. For neutrino-trapped matter, the expression becomes:
\begin{equation}
    sT=p_t+\varepsilon_t -n_B\mu_B-\mu_{\nu_e}(n_{\nu_e} +n_e),
\end{equation}
where $\mu_{\nu_e}$ is the neutrino chemical potential, and $n_{\nu_e}$ and $n_e$ are the number densities of electron neutrinos and electrons, respectively. To calculate the EoS together with the temperature fluctuations, we fix the entropy per baryon of the system, $s_B = \frac{s}{n_B}$, in units of the Boltzmann constant ($k_B$), which is set to unity in accordance with the natural unit convention.

\begin{table}[H]
\caption {DDME2 parameters.}
\begin{center}
\begin{tabular}{ | c | c | c | c | c | c | c | }
\hline
 meson($i$) & $m_i(\text{MeV})$ & $a_i$ & $b_i$ & $c_i$ & $d_i$ & $g_{i N} (n_0)$\\
 \hline
 $\sigma$ & 550.1238 & 1.3881 & 1.0943 & 1.7057 & 0.4421 & 10.5396 \\  
 $\omega$ & 783 & 1.3892 & 0.9240 & 1.4620 & 0.4775 & 13.0189  \\
 $\rho$ & 763 & 0.5647 & --- & --- & --- & 7.3672 \\
 \hline
\end{tabular}
\label{T}
\end{center}
\end{table}

\begin{table}
\caption {The ratio of the baryon coupling to the corresponding nucleon coupling for hyperons and $\Delta$s.}
\begin{center}
\begin{tabular}{ | c | c | c | c | c | } 
\hline
 b,d & $\chi_{\omega b,d}$ & $\chi_{\sigma b,d}$ & $\chi_{\rho b,d}$ & $\chi_{\phi b}$  \\
 \hline
 $\Lambda$ & 0.714 & {0.646} & 0 & -0.808  \\  
$\Sigma^0$ & 1 & {0.663} & 0 & -0.404  \\
  $\Sigma^{-}$, $\Sigma^{+}$ & 1 &  {0.663} & {1} & -0.404  \\
$\Xi^-$, $\Xi^0$  & 0.571 &{0.453} & 0 & {-1.01} \\
  $\Delta^-$, $\Delta^0$, $\Delta^+$, $\Delta^{++}$   & 1.285  & {1.331} & 1 & 0  \\
  \hline
\end{tabular}
\label{T1}
\end{center}
\end{table}

\section{stellar structure} \label{stellar}

{In this work the} PNSs are assumed to be axisymmetric stars with stationary rotation in which the matter can have only azimuthal motion, so that, we do not consider effects from convection, mass flows induced by heat transport, or meridional currents, for example. This way, the spacetime metric can be given by the following line element in quasi-isotropic spherical coordinates \cite{Paschalidis:2016vmz}:
\begin{equation}
ds^{2}=-e^{\gamma + \rho }dt^{2}+e^{2\alpha }\left( dr^{2}+r^{2}d\theta ^{2}\right)
+e^{\gamma - \rho}r^{2}\sin ^{2}\theta \left( d\phi -\omega dt\right) ^{2},
\label{metric}
\end{equation}
where $\alpha$, $\gamma$, $\rho$ and $\omega$\footnote{Note that $\rho$, $\omega$, $\gamma$, and $\phi$ defined in this section describe different physical quantities than those in Section \ref{mp}} are the metric functions which depend only on $r$ and $\theta$, and in this work we use natural units $(G=c=1)$. The function $\omega$ is associated with dragging the local inertial frame, the so-called Lense--Thirring effect \cite{pfister2007GReGr,ciufolini2004Natur}. We also consider that the matter that composes the star can be described by the energy-momentum tensor of a perfect fluid
\begin{equation}
T^{\mu\nu}=pg^{\mu\nu}+\left( \varepsilon +p\right) U^{\mu}U^{\nu},  \label{tmunu}
\end{equation}
where $\varepsilon$ and $p$ are the energy density and the pressure, respectively, $g^{\mu\nu}$ is the metric tensor and $U^{\mu}$ and $U^{\nu}$ are the four-velocity of the fluid. The conservation of the energy-momentum tensor, $\nabla_{\mu}T^{\mu\nu}=0$, when effects of temperature and stationary rotation are considered, will give \cite{bonazzola1993A&A,marques2017PhRvC}
\begin{equation}
    \partial_i \left( H + \ln \left[\frac{N}{\Gamma}\right] \right) = \frac{T e^{-H}}{m_B} \partial_i s - U_{\phi} U^t \partial_i \Omega, \label{hidroeq}
\end{equation}
where $H$ is the pseudo-enthalpy $H=\ln \left[(p+\varepsilon)/(m_B n_B)\right]$, $N=\sqrt{e^{\gamma + \rho }}$ is the lapse function, $\Gamma = N U^t$ is the Lorentz factor and $\Omega=U^{\phi}/U^t$ is the angular velocity.  {Since we are considering isentropic snapshots of the PNSs evolution $(\partial_i s=0)$ and that the stars are rigidly rotating $(\partial_i \Omega=0)$, the right side of Eq.~(\ref{hidroeq}) will vanish,} and this equation will reduce to the following
\begin{equation}
   \frac{\partial_i p}{\varepsilon+p} = \frac{\partial_i U^t}{U^t}. \label{hidrofinal}
\end{equation}
Therefore, combining the Einstein field equations, calculated using Eqs.~(\ref{metric}) and~(\ref{hidroeq}), with the hydrostatic equilibrium Eq.~(\ref{hidrofinal}), we obtain the set of equations that rules the stages of evolution of the PNSs, where an isentropic EoS describes each stage. We solve the equations with a numerical code based on the self-consistent field method proposed in~\cite{komatsu1989MNRAS,cook1994ApJ}.  It is important to mention that the EoS captures the local thermodynamic properties of matter, such as the pressure as a function of energy density. Therefore, according to the equivalence principle, spacetime always appears locally Minkowskian, even in a curved background. Rotational effects, such as frame-dragging, manifest in the global spacetime structure through modifications to the metric.

For cold, catalyzed, and static NS matter, the turning point theorem states that the onset of secular instability against axisymmetric perturbations occurs where $(dM/d\varepsilon_c)=0$, marking a transition from stable to unstable configurations. However, we need to extend the turning point criterion when considering hot and rigidly rotating PNSs as discussed in~\cite{1982ApJ...257..847S,friedman1988ApJ}. In this case, if we consider a set of equilibrium models parameterized by the central energy density $\varepsilon_c$, we need 3 out of 4 of these derivatives:
\begin{equation} 
    \frac{\partial M}{\partial \varepsilon_c},\quad \frac{\partial M_0}{\partial \varepsilon_c}, \quad\frac{\partial J}{\partial \varepsilon_c}, \quad \frac{\partial S}{\partial \varepsilon_c}, \label{eq_stab}
\end{equation}
where $M$ is the gravitational mass, $M_0$ is the baryonic mass, $J$ is the total angular momentum and $S$ is the total entropy, to vanish so that we have a turning point determining the transition from stable to unstable models~\cite{marques2017PhRvC,kaplan2014ApJ}. For sequences of isentropic PNSs $S=s_B M_0$~\cite{marques2017PhRvC} and we need to satisfy 2 out of the first 3 conditions of Eq.~(\ref{eq_stab}) if in addition, we consider sequences with constant angular momentum, then stability condition can be reduced to 
\begin{equation}
    \frac{\partial M}{\partial \varepsilon_c} \bigg|_{J,S} \geq 0.
\end{equation}
Another possible criterion is considered in~\cite{Goussard:1996dp}, $(\partial J/\partial n_B^c)|_{N,S}=0$, where $n_B^c$ is the central baryon density and $N$ is the  total baryon number. On the other hand, secular instabilities, such as r-mode and f-mode instabilities, operate on time scales from about $1$\,s to several years. These instabilities become significant at temperatures around $T \gtrsim 10^{11}$~K so that they typically occur after the first minute of the PNS lifetime~\cite{martinon2014PhRvD,andersson2001IJMPD,andersson2003CQGra}. In our results, we analyze the relation between non-axisymmetric instabilities and the ratio of rotational kinetic to gravitational binding energy $T/W$.

\section{Results and Analysis}
\label{results}

In all our figures, solid lines represent static stars, dashed lines represent stars rotating at $346.53$~Hz, which is the frequency of PSR J0740$+$6620 \cite{NANOGrav:2019jur}, dash-dot curves represent stars rotating at $716$~Hz, the frequency of PSR J1748$-$2446ad the fastest known NS \cite{hessels2006Sci}, and dash double-dot represent stars rotating at the {Kepler frequency, which is given by:
\begin{equation}
    \Omega_{K}=\left.\left(\omega+\frac{r\partial_{r}\omega}{2+r\partial_{r}\gamma-r\partial_{r}\rho}+\sqrt{\left(\frac{r\partial_{r}\omega}{2+r\partial_{r}\gamma-r\partial_{r}\rho}\right)^{2}+\frac{e^{2\rho}(\partial_{r}\gamma+\partial_{r}\rho)}{r(2+r\partial_{r}\gamma-r\partial_{r}\rho)}}\;\right)\right|_{r=R_{e},\theta=\frac{\pi}{2}},
\end{equation}
where $\rho$, $\gamma$ and $\omega$ are the metric functions from Eq.~(\ref{metric}). The Kepler frequency is reached when the equatorial angular velocity at the star's surface equals the angular velocity of a particle in a circular Keplerian orbit at the equator. At this point, the mass-shedding limit is attained, rendering the star unstable.

In this work we analyze four snapshots of the PNSs' evolution from birth to maturity as cold NSs, each snapshot represents a stage of evolution with constant values of entropy per baryon and lepton fraction ($Y_l$). The first stage ($s_B=1, Y_l=0.4$) represents the moment following core birth when the PNS has low entropy in its core and neutrinos are trapped due to high densities. Neutrino trapping heats the star, playing a critical role in stabilizing the PNS and creating the conditions necessary for the eventual explosion. Shortly after the core bounce ($\sim 0.5 - 1$~s), intense neutrino loss takes place decreasing the lepton fraction while increasing the entropy of the stellar matter and consequently the temperature. This defines the second stage considered ($s_B=2, Y_l=0.2$). The third stage ($s_B=2, Y_{\nu_e}=0$) occurs after some time of around $\sim 10-15$~s when the PNS completes deleptonization, becomes increasingly neutrino-transparent, and reaches its peak temperature. After this phase, the PNS starts cooling down until it reaches its final stage of evolution as a cold-catalyzed NS ($T=0$~MeV). A detailed discussion of the evolution of NSs from birth to maturity can be found in the following Refs.~\cite{Pons:2001ar, Janka:2012wk}.

\begin{figure}[!t]
    \includegraphics[width=0.49\textwidth]{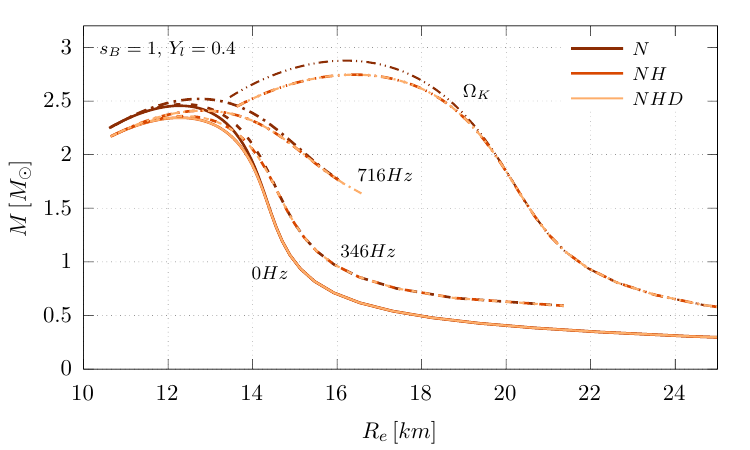}
    \includegraphics[width=0.49\textwidth]{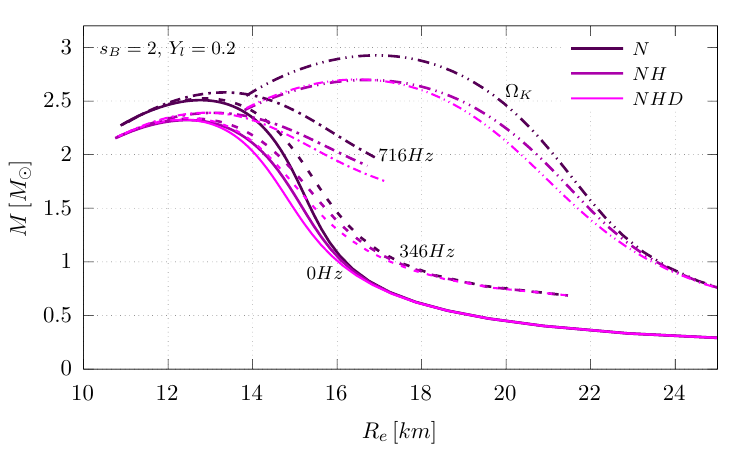}
    \includegraphics[width=0.49\textwidth]{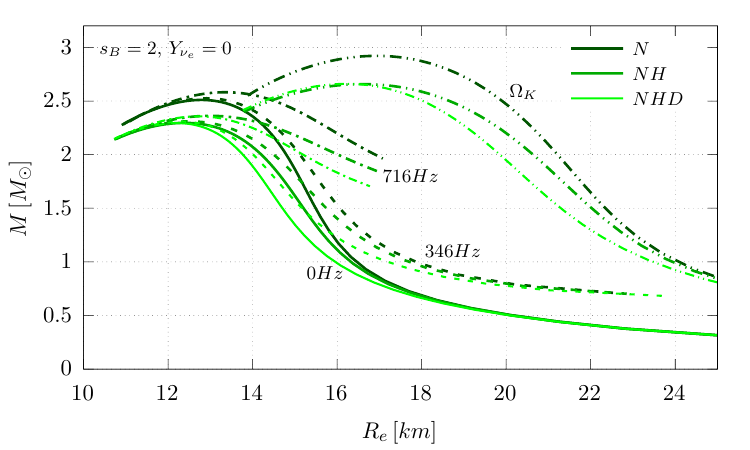}
    \includegraphics[width=0.49\textwidth]{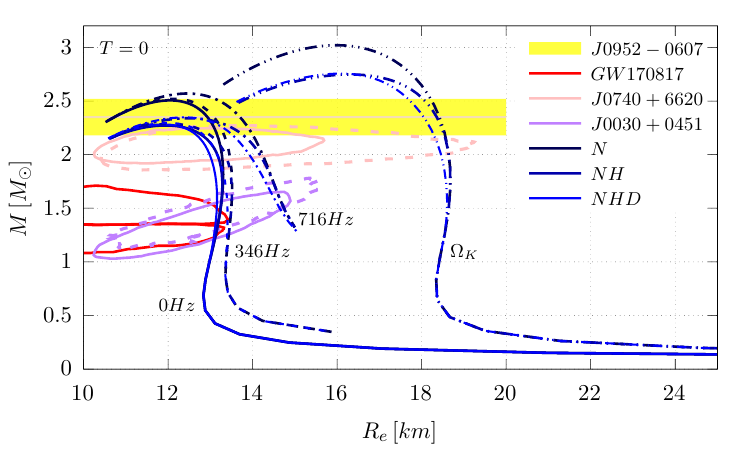}
    
    \caption{The gravitational mass $M$ as a function of the equatorial radius $R_e$ for four snapshots of the NS evolution. The two upper plots show the neutrino-trapped regime and the two lower plots show the neutrino-transparent regime. On the bottom right plot, for the cold NS, we also show the following constraints: purple contours for PSR J0030$+$0451 (solid line~\cite{riley2019} dashed line~\cite{Miller:2019cac}), pink contours for  PSR J0740$+$6620 (solid line~\cite{riley2021} dashed line~\cite{miller2021}), red contours for GW170817~\cite{abbott2019PhRvX} and yellow shaded region for PSR J0952$-$0607~\cite{bassa2017ApJ,romani2022ApJ}.}
    \label{figmxr}
\end{figure}

In Fig.~\ref{figmxr}, we show the relation between the gravitational mass and the equatorial radius, we can observe that for all stages of evolution analyzed the mass and the radius of the PNSs increase with the increasing of the rotation rate. For the lowest frequency considered ($346$~Hz), the $\rm M_{max}$ and the radius at $\rm M_{max}$ change very little at each stage, however, the radius of the stars at the intermediate to lower mass regions of the curve show a significant increase. For the higher rotation frequency of $716$~Hz, we observe a noticeable increase in the maximum mass and radius at each snapshot. Moreover, we observed that for the EoSs considered, only PNSs with $M \gtrsim 1.7$~\msun~can rotate with a frequency of $716$~Hz. In contrast, at the final stage, when the stars have cooled and become more compact, it is possible for NSs with $M \sim 1.4$~\msun~to rotate at that rate. When we consider the case of the Kepler frequency $\Omega_K$, we obtain a great increase in both mass and radius for all stars at all stages of evolution. 

For hot stars (upper panels and lower-left panels of Fig.~\ref{figmxr}), we observe an increase of approximately $1.18$ times in the value of $\rm M_{max}$ and about $1.32$ times in its radius, compared to the static configuration. Our result for the increase in $\rm M_{max}$ of the PNSs is slightly higher than the ones found in~\cite{khadkikar2021PhRvC} for stars with $s_B=3, Y_l=0.3$ but similar to what was determined in~\cite{nunna2020ApJ} for hot NSs, while the increase in the radius are in good agreement.

For cold stars (lower-right panel of Fig.~\ref{figmxr}), the maximum mass increases by $1.22$ times and the radius by $1.33-1.37$ times, both results agree with Ref.~\cite{khadkikar2021PhRvC}. For the cold pure nucleonic EoS we obtained a maximum mass of $3.02$~\msun~rotating with a Kepler frequency of $1527$~Hz, while for the EoS with hyperons and delta baryons, we obtain $\rm M_{max} = 2.76$~\msun~and $\Omega_K=1459$~Hz. For the first and the last stages of the evolution the presence of hyperons affects mainly the more massive stars reducing $\rm M_{max}$, while for the second and third stages, when the stars have higher entropy, the hyperons also affect the stars at the intermediate-mass region reducing their radii. As for the $\Delta$ baryons, they show almost no effect in the first snapshot. As the stars deleptonize and become hotter, it becomes noticeable that stars with intermediate mass have a smaller radius when $\Delta$'s are present. When the PNSs form cold-catalyzed NSs the effect of the exotic baryons persists but, its impact becomes less significant than for the stages with $s_B=2$. 

In general, we observe in Fig.~\ref{figmxr} that the difference in mass and radius between stars with only N and those with NH (nucleon plus hyperons) or NHD (nucleon plus hyperons plus $\Delta$-resonances) is slightly higher when we compare stars that are rotating than when they are static. The bottom right plot shows some observational constraints from NICER and gravitational wave data. The purple contours show the mass-radius estimates from NICER for the pulsar PSR J0030$+$0451, the solid contour shows the estimate obtained by Riley \textit{et al.}~\cite{riley2019}, $M=1.34_{-0.16}^{+0.15}$~\msun~and $12.71_{-1.19}^{+1.14}$~km bounded by $16\%$ and $84\%$ quartiles, and the dashed contour shows the estimate from Miller \textit{et al.}~\cite{Miller:2019cac}), $M=1.44_{-0.14}^{+0.15}$~\msun~and $13.02_{-1.06}^{+1.24}$~km at $68\%$ credibility. The pink contours show the mass-radius estimates from NICER for the massive pulsar PSR J0740$+$6620, and the solid contour shows the estimate obtained by Riley \textit{et al.}~\cite{riley2021}, $M=2.072_{-0.066}^{+0.067}$~\msun~and $R=12.39_{-0.98}^{+1.30}$~km bounded by $16\%$ and $84\%$ quartiles, and the dashed contour shows the estimate from Miller \textit{et al.}~\cite{miller2021}), $R = 13.71^{+2.61}_{-1.5}\rm km$ at $68\%$ credibility. The red contours show the mass and radius encountered by the Virgo/LIGO collaboration for the stars involved in the GW170817 event~\cite{abbott2019PhRvX,abbott2019PhRvX}, $M_1 = 1.46^{+0.12}_{-0.10}$~\msun~and radius $R_1 = 10.8^{+2.0}_{-1.7}$~km and $M_2 = 1.27^{+0.09}_{-0.09}$~\msun~and radius R$_2 = 10.7^{+2.1}_{-1.5}$~km with $90\%$ credible intervals. The yellow shaded region shows the estimated mass of PSR J0952$-$0607~\cite{bassa2017ApJ,romani2022ApJ}, $2.35 \pm 0.17$~\msun, which is the fastest ($707$~Hz) known NS in our galaxy. We can observe that the mass-radius curves for static NSs satisfy all the constraints considered, the curves for $346$~Hz satisfy the constraints from NICER and  PSR J0952$-$0607 with larger radii than the static configuration but, for the EoSs considered in this work, the NSs in the GW170817 event are likely rotating with $\Omega < 346$~Hz. Our results also show that it is possible to have NSs with approximately the same mass-radius constraints as the ones from NICER rotating with a fast rotation of $716$~Hz. We also obtain that if PSR J0740$+$6620 were rotating at the Kepler limit, it would have a minimum radius of $18.5$~km with the NHD EoS.

\begin{figure}[!t]
   \includegraphics[width=0.49\textwidth]{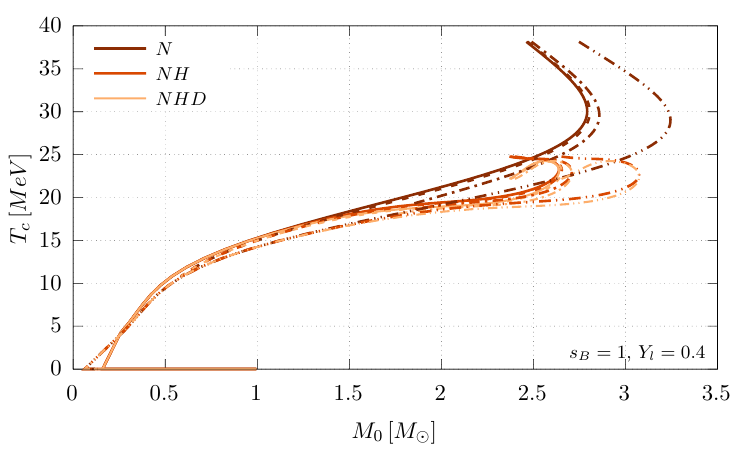}
    \includegraphics[width=0.49\textwidth]{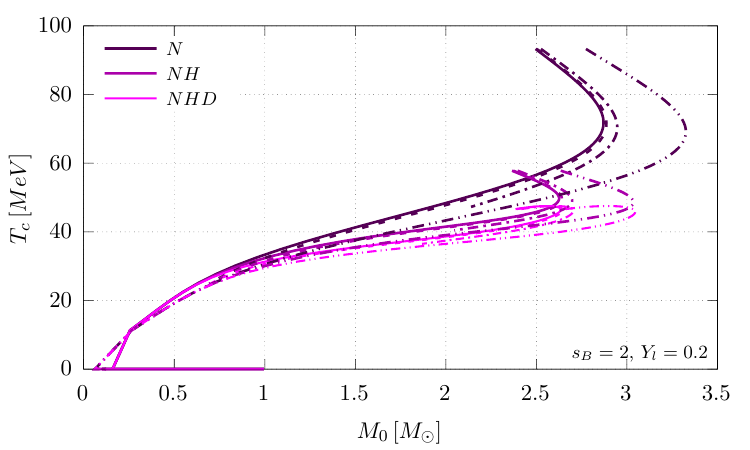}
    \includegraphics[width=0.49\textwidth]{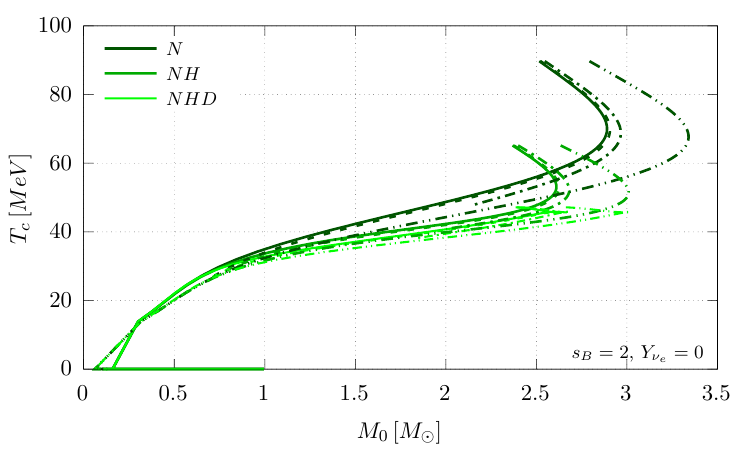}
       
    \caption{The central temperature as a function of the baryonic mass for three snapshots of the NS evolution. The two upper plots show the neutrino-trapped regime and the lower plot shows the neutrino-transparent regime. In all plots solid lines represent static stars, dashed lines represent stars rotating at $346.53$~Hz, dash-dot curves represent stars rotating at $716$~Hz, and dash double-dot represent stars rotating at $\Omega_K$. Notice that the top left plot has a different temperature scale.}
    \label{figm0xt}
\end{figure}

In Fig.~\ref{figm0xt}, we show the plot of the central temperature against the baryon mass at the first three stages of the stellar evolution considered, {namely, the upper-left panel corresponds to the initial stage with small entropy density, $s_B=1$, and large lepton fraction $Y_l=0.4$, the upper-right panel shows the deleptonization stage with $s_B=2$ and $Y_l=0.2$, and the lower panel presents neutrino transparent stage corresponding to $s_B =2$ and $Y_{\nu_e}=0$}. The baryon mass is a quantity that is expected to remain conserved throughout the stellar evolution unless there is mass lost from winds, tidal trapping, or mass gain through accretion, phase transformation, or gravitational collapse.  The gravitational mass is associated to the gravitational field generated by the star, and it is the mass that is determined by observational constraints; this way, it is more interesting to use the gravitational mass in Fig. \ref{figmxr}. The baryonic mass, it is associated to the total number of baryons inside the star, and it is assumed to be conserved as the PNS evolves, this way, if we plot the temperature as a function of the baryonic mass we can follow the evolution of specific stars by looking to fixed values of $M_0$. 

The current study assumes hydrostatic equilibrium of the stellar matter where the baryon mass is conserved through stellar evolution. On the other hand, the central temperature of the PNSs is higher after the core bounce and starts decreasing due to neutrino cooling and thermal radiation after deleptonization. Rotation introduces centrifugal forces that change the equilibrium structure of the matter, thereby impacting temperature distributions and changes in the maximum mass limit. The centrifugal force decreases the effective gravitational pull, allowing the star to support more mass at a given temperature. 

From Fig. \ref{figm0xt}, we observe that rotating stars have lower $T_c$ compared to the static ones. Moreover, the higher the rotating frequency, the lesser the $T_c$ of the PNSs. Aside from that, increasing baryon degrees of freedom tends to lower $T_c$, as clearly shown in all three panels, {independent on the lepton fraction. Along the star evolution following upper left (first stage), right (second stage), and the bottom panel (third stage),} the $T_c$ is lower when the star is young and neutrino-rich. After the core bounce, neutrino diffusion together with external shocks increases $T_c$. 

In the third stage, shown in the lower panel of Fig.~\ref{figm0xt}, when the neutrinos have escaped from the stellar core, the temperature reaches its peak and begins to drop through neutrino cooling and thermal radiation. The decrease in $T_c$ due to cooling processes reduces the thermal support in the PNSs, leading to contraction from diminished thermal pressure and potentially lowering the baryonic mass threshold for stability~\cite{Koliogiannis:2020nhh, Lenka:2018ehb}.  Notice that, similar to the mass-radius relation, the behavior of $T_c$ as a function of $M_0$ is reversed after reaching the maximum baryonic mass ($M_0^{\rm max}$). From low masses up to $M_0^{\rm max}$, $T_c$ increases with increasing $M_0$. However, beyond this point, 
$T_c$ increases as $M_0$ decreases, which clearly indicates that the compression resulting from the decrease in the star's size, as seen in Fig.~\ref{figmxr}, drives the rise in temperature. Additionally, the other relations analyzed in this work also present a similar pattern.

\begin{figure}[!t]
    \centering
    \includegraphics[width=0.49\textwidth]{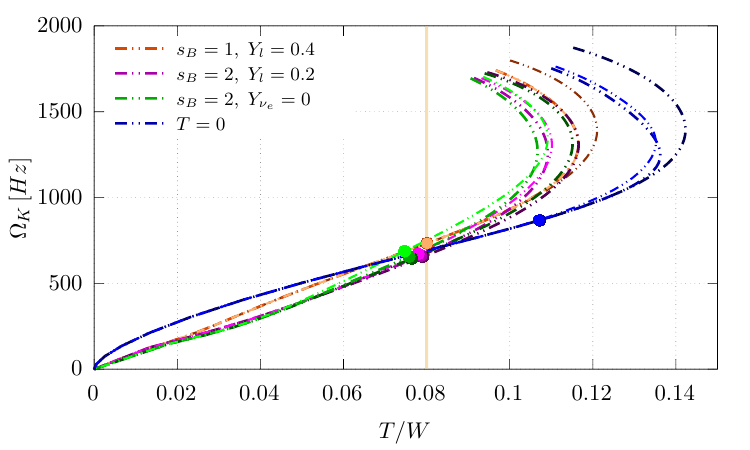}
    \includegraphics[width=0.49\textwidth]{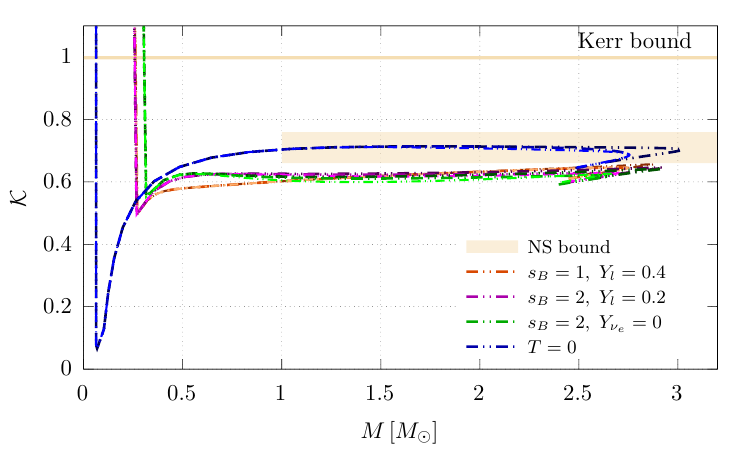}
    \caption{On the left we show the Kepler frequency as a function of the ratio of rotational kinetic to gravitational binding energy. The vertical line represents the critical value $T/W = 0.08$ for gravitational radiation instabilities, and the position of the $\rm M=1.4$~\msun~is highlighted by discs. On the right, we show the Kerr parameter as a function of gravitational mass for stars at the mass-shedding limit. The horizontal line indicates the Kerr limit for astrophysical Kerr BHs, $\mathcal{K}_{\text{BH}} = 0.998$~\cite{thorne1974}. The shaded area corresponds to the constraints for NSs obtained from Eq.~(\ref{eqkerrk}).}
    \label{figok}
\end{figure}

On the left side of Fig.~\ref{figok}, we show the Kepler frequency $\Omega_K$ as a function of the ratio of the rotational kinetic energy to gravitational binding energy $T/W$ for the four snapshots of the PNSs evolution. The NSs can emit gravitational waves due to no-axisymmetric perturbations and the ratio between the kinetic energy $T$ and the gravitational potential energy $W$ can provide us a way to detect the point where instabilities driven by gravitational radiation would begin. Morsink~\textit{et al.}~\cite{morsink1999ApJ} considered a wide range of realistic EoSs and established that for stars with gravitational mass of $1.4$~\msun~the instabilities set in at $T/W \sim 0.08$, which is shown in our figure by a vertical line. We can observe in this figure that for the EoSs considered in this work if a star of mass $1.4$~\msun~is born rotating at the Kepler limit it will be approximately inside the limit for non-axisymmetric instability. As the $1.4$~\msun~PNS starts to deleptonize and become hotter, the ratio $T/W$ decreases and it becomes more unlikely that the instabilities occur. However, when the stars become cold and catalyzed the ratio of the rotational kinetic to gravitational binding energy increases greatly, since the stars become more compact, so at this stage, the instabilities driven by gravitational radiation will begin before the stars reach the Kepler limit. In this case, the star will lose angular momentum through the emission of gravitational waves until the star becomes stable~\cite{morsink1999ApJ}. This way, our results indicate that although a canonical NS can be born maximally rotating, it is likely to spin down as it cools due to the loss of angular momentum by gravitational radiation. 

An additional quantity that we can investigate when studying rotating PNSs is the the Kerr parameter $\mathcal{K}$, on the right side of Fig.~\ref{figok}, we display $\mathcal{K}$ as a function of the gravitational mass $M$ for stars at the Kepler limit for the four stages of evolution. The Kerr parameter, or dimensionless spin parameter, is given by
\begin{equation}
    \mathcal{K} = \frac{J}{M^2} =  \frac{I \Omega}{M^2}, \label{eqkerr}
\end{equation}
where $J$ is the total angular momentum, $\Omega$ is the rotation frequency and $I$ is the moment of inertia around the rotation axis, for a rigidly rotating star $I=J/\Omega$. The Kerr parameter is an important physical quantity related to BHs and to the final fate of rotating NSs, specially the ones rotating at the mass-shedding limit. When a massive rotating NS collapses, due to the conservation of angular momentum and mass-energy, the BH that is generated has almost the same mass and angular momentum of the originating NS, and consequently almost the same $\mathcal{K}$. Kerr BHs have a maximum value of $\mathcal{K}_{\text{BH}} = 0.998$~\cite{thorne1974}, indicated by a horizontal line in our figure, while for cold NSs, a maximum value of $\mathcal{K}_{\text{NS}} \sim 0.75$ was obtained in~\cite{koliogiannis2020PhRvC}. Besides, Koliogiannis \& Moustakidis~\cite{Koliogiannis:2020nhh} obtained a simple universal relation between $\mathcal{K}$ and the compactness of the maximum mass configuration for cold, catalyzed NSs given by
\begin{equation}
    \mathcal{K}_K \simeq 1.34 \sqrt{ \frac{\rm M_{max}^{rot}}{\left. R_{e} \right. _{\rm max}^{\rm rot}}} \label{eqkerrk},
\end{equation}
where $\rm M_{max}^{rot}$ is the maximum mass of the rotating cold, catalyzed NS and $\left. R_{e} \right. _{\rm max}^{\rm rot}$ is the equatorial radius for the same configuration. The relation of Eq.~(\ref{eqkerr}) is shown as a shaded area in our figure and indicates that for cold NSs with $M>1$ \msun~and rotating at $\Omega_K$ the value of $\mathcal{K}$ is almost independent of the EoS. We can observe that for the EoSs considered in this work, the ``hot'' NSs also follow an almost universal relation for $M>1$ \msun, but on a lower range of values for $\mathcal{K}$. For the EoSs considered in this work, we encountered a maximum value of $\mathcal{K} \sim 0.71$ for cold NSs with nucleonic EoSs. This is consistent with the results found in~\cite{koliogiannis2020PhRvC}, which indicate that stiffer EoSs lead to higher maximum values of the Kerr parameter in NSs. In the first evolutionary stage, the Kerr parameter increases with the increase of the gravitational mass, in the second and third stages, the $\mathcal{K}$ values decrease with $M$ until around $M \sim 1.5$~\msun~and then start to increase. As mentioned above, for cold NSs, we observe that for $M \gtrsim 1~\mathrm{M_\odot}$, the Kerr parameter remains nearly constant. However, taking a closer look at it, we can observe that for hyperonic NSs the value of $\mathcal{K}$ starts to slightly decrease for $M \gtrsim 2$~\msun, while for NSs with $\Delta$'s the decrease starts in $M \gtrsim 1.25$~\msun, which indicates that the inclusion of new degrees of freedom can make the Kerr parameter decrease with the increasing of the mass, for masses above a certain value. In general, we observe that the inclusion of hyperons leads to a decrease in the value of $\mathcal{K}$, and the inclusion of $\Delta$'s leads to further decrease, in all stages of evolution. Additionally, the increase in the temperature of the PNS, and as a consequence, the softening of the EoS, also lead to a decrease in the Kerr parameter.

\begin{figure}[!t]
    \includegraphics[width=0.49\textwidth]{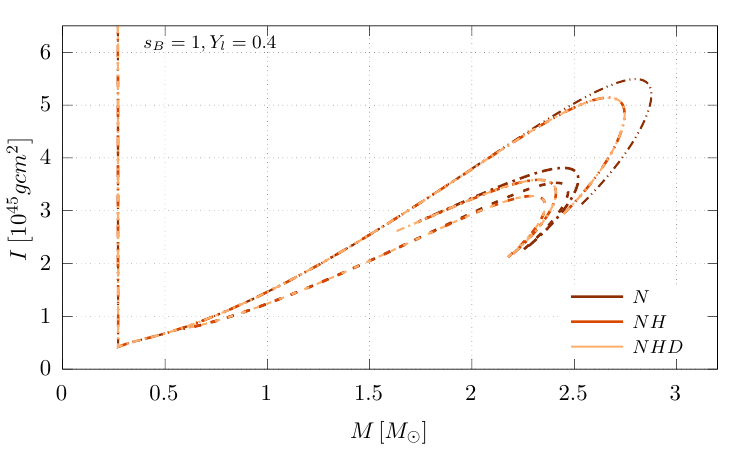}
    \includegraphics[width=0.49\textwidth]{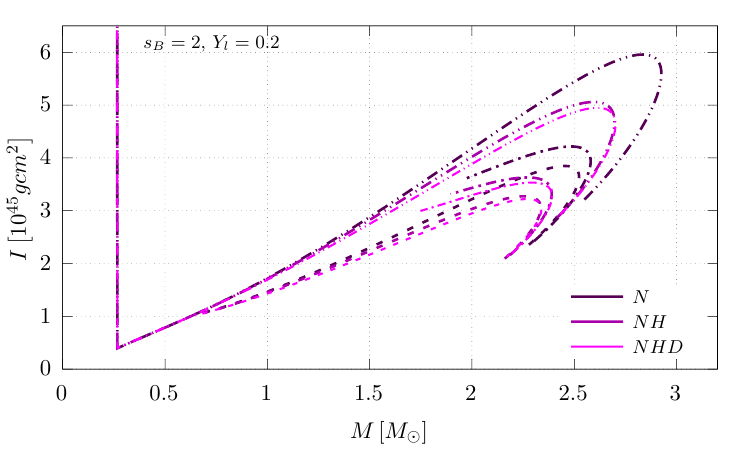}
    \includegraphics[width=0.49\textwidth]{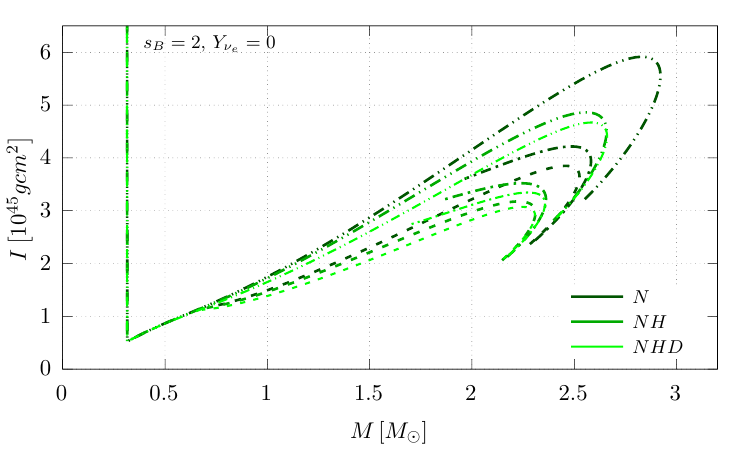}
    \includegraphics[width=0.49\textwidth]{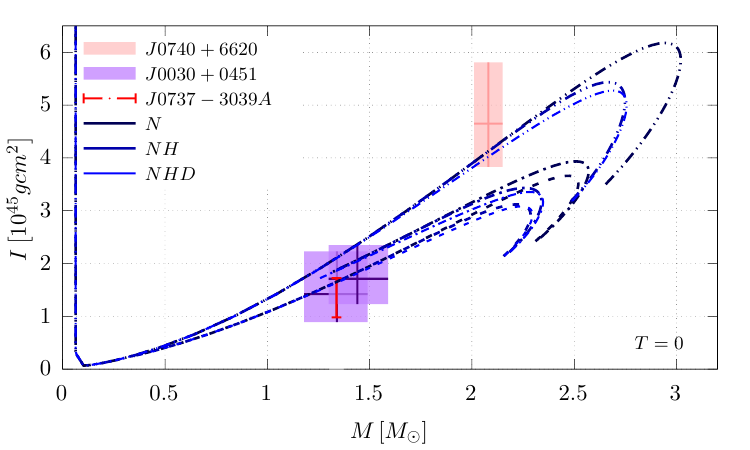}
    \caption{The moment of inertia as a function of the gravitational mass for four snapshots of the NS evolution. The two upper plots show the neutrino-trapped regime, and the two lower plots show the neutrino-transparent regime. On the bottom right plot, for the cold NSs, we also show the following constraints: purple shaded regions for PSR J0030$+$0451~\cite{silva2021PhRvL}, pink shaded region for PSR J0740$+$6620~\cite{li2022CQGra} and red bar for PSR J0737$-$3039A~\cite{bejger2005MNRAS}. In all plots dashed lines represent stars rotating at $346.53$~Hz, dash-dot curves represent stars rotating at $716$~Hz, and dash double-dot represent stars rotating at $\Omega_K$.
    }
    \label{figmxi}
\end{figure}

In Fig.~\ref{figmxi} we show the moment of inertia $(I)$ as a function of the gravitational mass. The moment of inertia is associated with the distribution of mass relative to the axis of rotation of the star and it measures the star's resistance to changes in its rotational motion, for a uniformly rotating star $I = 2 \pi J/\Omega$, where $J$ is the angular momentum and $\Omega$ is the angular frequency. On the other hand, the gravitational mass represents the star's total mass, which determines its gravitational influence and dictates the strength of its gravitational field. These two physically independent quantities, the moment of inertia and gravitational mass, are connected through rotational dynamics, as both influence the rotational behavior of a star. For a spherically symmetric, rigidly rotating star with a uniformly distributed mass, the moment of inertia is given by $I = 2/5 MR^2$. For highly compact objects (lower radius and constant mass), the moment of inertia $I$ becomes smaller, making the star less resistant to rotational changes. 

Comparing the snapshots in Fig.~\ref{figmxi}, we observe that, for stars rotating at the Keplerian frequency, $I$ increases from the first panel (representing the birth of the star) and reaches its maximum when the star becomes cold and catalyzed. On the other hand, for stars rotating at $346$~Hz and $716$~Hz, the value of $I$ increases as the PNSs heats up and then decreases again when the stars become cold and catalyzed. This suggests that $I$ can change in different ways as the PNSs evolve if they are rotating at the maximum rotational rate or below this limit. We can observe from the snapshots that the emergence of heavy baryons reduces $I$, as the star becomes more compact. The appearance of these new degrees of freedom redistributes the available energy among a larger number of particle species, lowering the average kinetic energy and Fermi momentum. Consequently, the pressure support against gravitational collapse decreases, leading to a more compact configuration. This increased compactness, in turn, reduces the moment of inertia, as more mass is concentrated toward the stellar core. This implies that stars with a larger number of degrees of freedom are more susceptible to changes in rotational speed. Additionally, in the bottom right plot we also show the inferred estimates of $I$ for PSR J0740$+$6620~\cite{li2022CQGra}, PSR J0030$+$0451~\cite{silva2021PhRvL}, and the active radio pulsar PSR J0737$-$3039A~\cite{bejger2005MNRAS}, for comparison. We emphasize that the binary NS system PSR J0737$-$3039 is highly relativistic~\cite{burgay2003Natur,kramer2009CQGra}, such that, it is an important system to execute tests of general relativity and modified theories of gravity. In particular, it is believed that in the future it will be possible to measure the moment of inertia of the pulsar PSR J0737$-$3039A. So, several works in the literature are directed towards estimating the moment of inertia of this NSs, see for example~\cite{morrison2004ApJ,bejger2005MNRAS,lattimer2005ApJ,landry2018ApJ,lim2019PhRvC,miao2022MNRAS}. In Fig.~\ref{figmxi}, we can observe that our results are in agreement with the constraints for PSR J0030$+$0451 and PSR J0737$-$3039A. However, only the curves for stars at the mass shedding limit satisfy the constraint for PSR J0740$+$6620. Our results predict that an NS with $\sim 2$~\msun~and rotating at $346$~Hz has a moment of inertia of $\rm \sim 3 \times 10^{45}~g cm^2$.

\begin{figure}[!t]
    \includegraphics[width=0.49\textwidth]{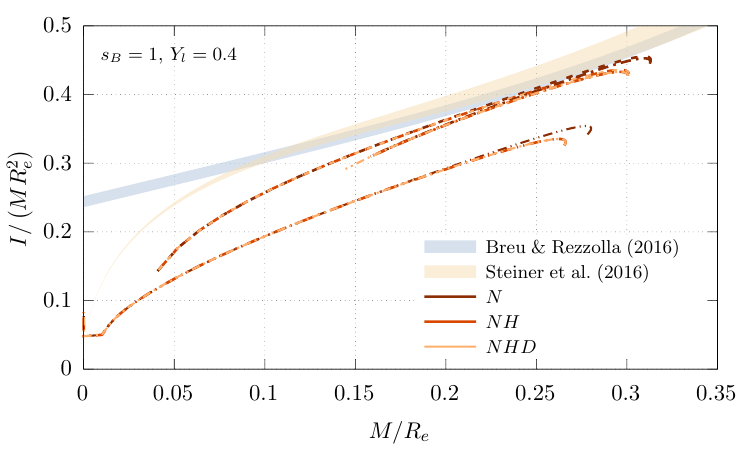}
    \includegraphics[width=0.49\textwidth]{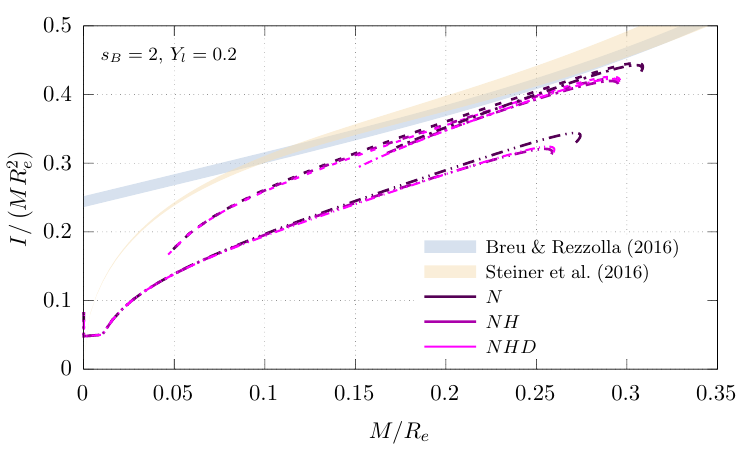}
    \includegraphics[width=0.49\textwidth]{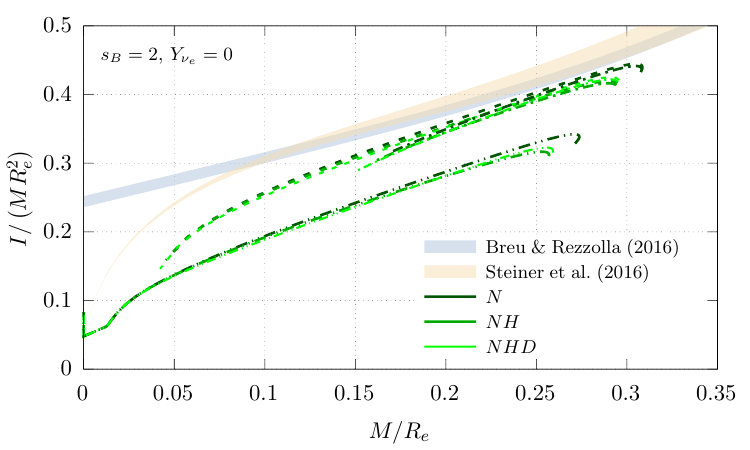}
    \includegraphics[width=0.49\textwidth]{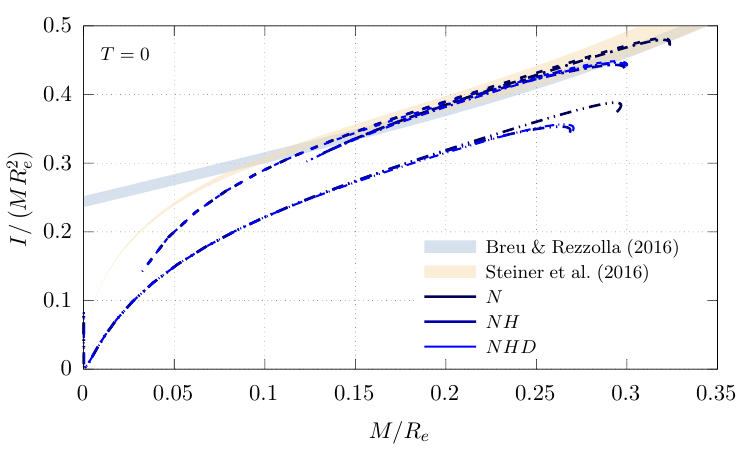}
    \caption{Dimensionless moment of inertia $I/(M R^2)$ as a function of compactness $M/R$ for four snapshots of the NS evolution. The two upper plots show the neutrino-trapped regime and the two lower plots show the neutrino-transparent regime. The correlation bands found by Steiner \textit{et al.}~\cite{steiner2015} are in gray, and the ones found by Breu \& Rezzolla~\cite{breu2016} are in beige. In all plots dashed lines represent stars rotating at $346.53$~Hz, dash-dot curves represent stars rotating at $716$~Hz, and dash double-dot represent stars rotating at $\Omega_K$.}
    \label{figixc}
\end{figure}

In Fig.~\ref{figixc}, we display the dimensionless moment of inertia $I/(M R_e^2)$ as a function of the compactness $M/R_e$ for four stages of evolution of the PNS. The dimensionless moment of inertia places a significant constraint on the internal structure of the NSs, besides, several studies~\cite{ravenhall1994ApJ,lattimer2001ApJ,lattimer2005ApJ,steiner2015,breu2016,greif2020ApJ} have shown that this quantity is strongly correlated with the compactness. In our figures, we compare our results with the correlation bands encountered by Steiner \textit{et al.}~\cite{steiner2015} and the ones found by Breu \& Rezzolla~\cite{breu2016}, which were obtained for cold NSs. In~\cite{Raduta:2020fdn} the authors proposed a corrected fit for PNSs with $s_B=2, Y_l=0.2$. Our results for cold NSs only agree with these bands for stars rotating below the Kepler limit and for $M/R_e \gtrsim 0.15$, a similar result was obtained in~\cite{greif2020ApJ}. In all stages of evolution, the correlations for stars at the mass shedding limit stay far below the curves for stars rotating at $346$ and $716$~Hz. In general, we observe that the correlations for the hot stars also fall below the bands encountered for the cold stars. This is because hotter, more rapidly rotating stars are less compact. 

\begin{table}[!ht]
 \caption{Summary of the physical properties of the PNSs at the mass shedding limit. All properties are related to the star with maximum gravitational mass $\rm M_{max}$ for each sequence. Where $\Omega_K$ is the Kepler frequency, $\mathcal{K}$ is the Kerr parameter, $n_c/n_0$ is the ratio of the central baryon density to the saturation density,  $T_c$ is the central temperature, $R_p/R_e$ is the ration of the polar radius to the equatorial radius, $T/W$ is the ratio of the rotational kinetic energy to the gravitational binding energy, $I$ is the moment of inertia, and $M/R_e$ is the compactness.}
    \centering
    \begin{ruledtabular}
      \setlength\extrarowheight{2pt}
    \begin{tabular}{ c c c c c c c c c c c c c c }

         $s_B,Y_l$ & EoS & $\rm M_{max}$ & $\rm M_0$ & $\rm R_e$ & $\rm n_c/n_0$ & $\rm T_c$ & $\rm J$ & $\Omega_K$ & $\rm \mathcal{K}$ & $\rm I [10^{45} $ & $\rm T/W$ & $\rm R_p / R_e$ & $\rm M / R_e$ \\
         
         &  & $[$\msun$]$ & $[$\msun$]$ & $\rm [km]$ &   & $\rm [MeV]$ & $[G M_{\odot}^2 / c]$ & $\rm [Hz]$ &  & $\rm g\,cm^2]$ & \\
         
         \hline
                & \ft N & 2.88 & 3.24 & 16.2 & 4.86 & 29.1 & 5.43 & 1467 & 0.656 & 5.19 & 0.120 & 0.57 & 0.263  \\

        $1, 0.4$  & \ft NH & 2.75 & 3.07 & 16.5 & 4.86 & 22.6 & 4.86 & 1403 & 0.644 & 4.85 & 0.116 & 0.57 & 0.246 \\

                & \ft NHD & 2.75 & 3.07 & 16.5 & 4.86 & 22.4 & 4.86 & 1403 & 0.644 & 4.85 & 0.116 & 0.57 & 0.246  \\

        \hline 

                & \ft N & 2.93 & 3.33 & 16.9 & 4.60 & 69.3 & 5.53 & 1385 & 0.646 & 5.59 & 0.116 & 0.57 & 0.255   \\

        $2, 0.2$  & \ft NH & 2.70 & 3.04 & 16.7 & 4.99 & 48.7 & 4.55 & 1362 & 0.626 & 4.68 & 0.108 & 0.58 & 0.239 \\

                & \ft NHD & 2.70 & 3.05 & 16.4 & 5.11 & 45.9 & 4.59 & 1394 & 0.629 & 4.61 & 0.109 & 0.58 & 0.243  \\
        
        \hline

                & \ft N & 2.92 & 3.34 & 17.0 & 4.60 & 67.6 & 5.48 & 1377 & 0.642 & 5.58 & 0.115 & 0.58 & 0.254\\

        $2, 0$  & \ft NH & 2.66 & 3.01 & 16.6 & 5.11 & 51.5 & 4.38 & 1362 & 0.620 & 4.50 & 0.106 & 0.58 & 0.236 \\

                & \ft NHD & 2.66 & 3.02 & 16.2 & 5.24 & 45.8 & 4.43 & 1408 & 0.626 & 4.41 & 0.109 & 0.58 & 0.242  \\
        
        \hline

              & \ft N & 3.02 & 3.57 & 16.0 & 4.77 & ... & 6.42 & 1527 & 0.704 & 5.89 & 0.141 & 0.56 & 0.278 \\

        $T=0$  & \ft NH  & 2.75 & 3.20 & 16.3 & 4.90 & ... & 5.19 & 1426 & 0.687 & 5.09 & 0.132 & 0.56 & 0.248  \\

            & \ft NHD  & 2.76 & 3.21 & 16.1 & 5.03 & ... & 5.22 & 1459 & 0.687 & 5.01 & 0.132 & 0.56 & 0.253 \\
    \end{tabular}
    \end{ruledtabular}
    \label{tab3}
\end{table}

In Tab.~\ref{tab3} we summarize some of the physical properties associated with the configuration with maximum gravitational mass and uniformly rotating at the mass shedding limit for each of the stages of evolution of the PNSs and for the three types of EoS considered: only nucleons (N), nucleons plus hyperons (NH) and nucleons plus hyperons plus $\Delta$s (NHD). As expected, EoS with nucleons only produce the highest values of $M_{max}$ and $M_0$ compared to EoS with NH and NHD. We also observe that the inclusion of $\Delta$ baryons has almost no effect on the maximum mass but leads to a slightly smaller radius than the EoS NH. As a consequence, the stars with NHD are more compact ($M / R_e$) than those with NH. In general, the EoS for nucleonic matter leads to the highest compactness. The Kepler frequency $\Omega_K$, the angular momentum $J$, the Kerr parameter $\rm \mathcal{K}$, and the ratio of rotational kinetic to gravitational binding energy $T/W$ are all proportional to the compactness. On the other hand, the central number baryon density ($n_c/n_0$) increases when hyperons are included and increases even further when $\Delta$s are added. Nonetheless, the central temperature $\rm T_c$ and the moment of inertia $I$ go in the opposite direction; that is, they decrease with NH and decrease again by adding $\Delta$s. In general, we observe that $\rm T_c$ seems to be the quantity that is most sensitive to the presence of $\Delta$-resonances, while the ratio between the polar and the equatorial radius $R_p / R_e$ seems to be the property least affected by the three types of EoS. As already observed in our figures, in the first stage of evolution considered ($s_B=1, Y_l=0.4$), we do not observe significant differences in the characteristics of the star with $M_{max}$ when we compare the EoS for NH and NHD. The stage when the temperature in the star has reached its peak ($s_B=2, Y_{\nu_e}=0$) is when the inclusion of $\Delta$ resonances leads to more noticeable effects. Additionally, we observe that the maximum baryonic mass for the EoS with only nucleons (N) increases throughout the evolutionary stages. However, this only can happen if we consider some type of mass accretion, otherwise, if $M_0$ is conserved, the maximum baryonic mass at each stage cannot be higher than $3.24$ \msun~(maximum $M_0$ in the first stage), so that, in the last stage we would have $\rm M_{max} = 2.79$~\msun~for the nucleonic EoS. The EoSs with hyperons and $\Delta$ baryons have the smallest $M_0$ maximum in the third stage, $3.01$ \msun~and $3.02$ \msun~respectively, causing PNSs with greater $M_0$ in the first two stages to collapse. For the NH and NHD EoSs, if the baryonic mass is conserved, the $\rm M_{max}$ at the last stage will be $2.62$ \msun~ for both EoSs.

\begin{figure}[!t]
    \includegraphics[width=0.49\textwidth]{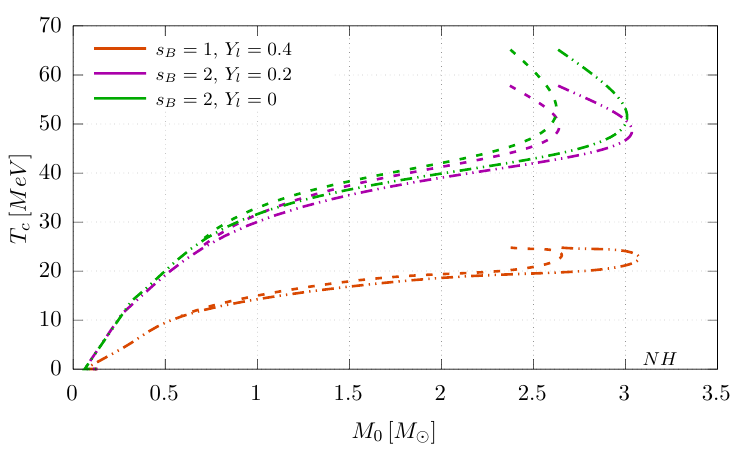}
    \includegraphics[width=0.49\textwidth]{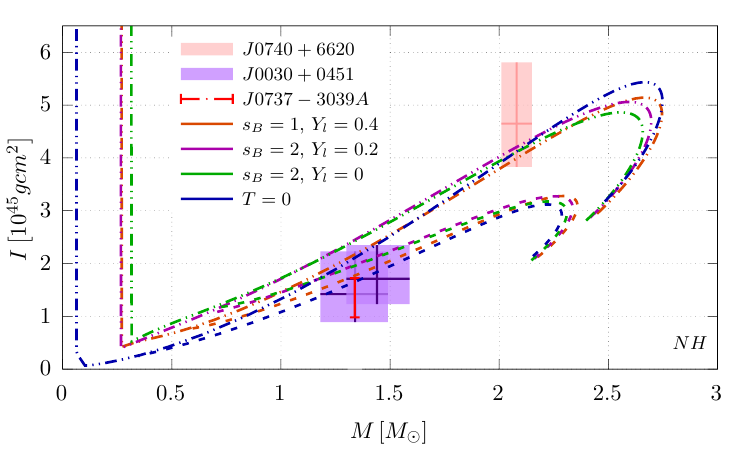}
    \caption{Extra plots: Dashed lines represent stars rotating at $346.53$~Hz and dash double-dot represent stars rotating at $\Omega_K$. }
    \label{figext}
\end{figure}

In Fig. \ref{figext}, we show two extra plots: the central temperature as a function of the baryonic mass, on the left, and the moment of inertia as a function of the gravitational mass, on the right. In this figure, we show the results for all stages of evolution in the same plot, but only for the NH EOS. This way, we can more easily observe how these quantities change as the PNS evolve. We summarize our findings for the rotating stars in Tab.~\ref{tab3}, where it is shown the their properties with $\rm M_{max}$ rotating at Kepler frequency. It is believed that NSs can born rotating at or very close to $\Omega_K$, however, at the moment, we have not detected any NS rotating at this limit. It is thought that this occurs because as the PNSs evolve, various mechanisms, such as magnetic winds and neutrino emission, can lead to significant loss of angular momentum, causing the rotation to slow down.

\section{Final Remarks and Conclusion}\label{remarks}

This work investigates the effects of rotation on the evolution of the macroscopic structure of PNSs, focusing on rigidly rotating PNSs with hyperons and $\Delta$-resonances in its core. We considered three different rotational frequencies: $346.53$~Hz, $716$~Hz, and the Kepler frequency $\Omega_K$ to analyze their effect on the star's structure. Three stellar matter configurations were considered: only nucleons, nucleons plus hyperons, and nucleons plus hyperon plus $\Delta$-resonances. This enables us to analyze the stellar composition \textit{vis-à-vis} the structural changes induced by rotation. Generally, the centrifugal force generated by the star's rotation reduces the effective gravitational pull, enabling the star to support a higher mass before collapsing. We observe that an increase in rotational frequency significantly impacts the star's radius in the intermediate region ($M<2$~\msun), with only a small difference in the maximum mass, particularly between $346.53$~Hz and $716$~Hz. However, at the mass-shedding limit, the maximum mass and radius exhibit significant differences compared to the static case. As expected, an increase in the baryonic degrees of freedom of stellar matter results in a reduction in both the maximum mass and radius. In particular, the main effect of the inclusion of $\Delta$ resonances on the mass-radius relation is to reduce the radius of stars with intermediate mass especially when the stars are hotter. We also compared our results for cold, catalyzed, and rigidly rotating NSs with the constraints for PSR J0740$+$6620 and PSR J0030$+$0451, GW170817 and PSR J0952$-$0607 and concluded that considering the EoSs used in this work the NSs involved in the GW170817 event likely rotate at frequencies lower than $346$~Hz. On the other hand, we observed that it is possible to have stars rotating at the mass shedding limit and obeying the constraints from~\cite{miller2021} for PSR J0740$+$6620.

We studied the conserved baryon mass against the central temperature and observed that $T_c$ is lower in a neutrino-rich stellar matter when the star is still young and increases during neutrino diffusion. After the neutrinos have escaped from the stellar core the $T_c$ reaches its peak and starts to decrease. This causes contraction and leads to a reduction in the baryon mass threshold for stability. We observed that rotating PNSs have lower central temperatures compared to static ones. The higher the rotation frequency, the lower the central temperature. Furthermore, when new baryonic degrees of freedom are introduced, it generally leads to a drop in the $T_c$ and also reduces the baryon mass threshold for stability, as the star contracts. 

From the variation of $\Omega_K$ against $T/W$, we observe that newly born stars can maintain stability while rotating rapidly. However, as they evolve, they may lose angular momentum through gravitational wave emission, leading to a significant reduction in their rotation rate. On the contrary, once instability sets in, the star may collapse gravitationally. Analyzing the $\rm M$ versus $\mathcal{K}$ relation, we observe that the emergence of hyperons and $\Delta$-baryons in the stellar core reduces the value of $\mathcal{K}$. This reduction is attributed to the softening of EoS, which lowers the maximum angular momentum and mass that the star can sustain. Additionally, the presence of heavy baryons decreases the rotational support of the star and reduces the maximum mass limit, potentially increasing the likelihood of BH formation during its evolution, even though it poses a lower $\mathcal{K}$ value.

The moment of inertia was calculated and studied in terms of compactness and gravitational mass. The results were compared to inferred values in the literature. Generally, a higher value of $I$ stabilizes the star against rapid changes in the rotational regime and reduces the efficiency of rotational energy loss. Our results showed that when new baryonic degrees of freedom appear in the stellar matter, $I$ decreases. This implies that stars with hyperons and $\Delta$-resonances in their cores are more responsive to rotational dynamics. We also observed that the moment of inertia behaves differently based on the star's rotation rate. For stars rotating at $\Omega_K$, $I$ increases and peaks when the star becomes cold and catalyzed. For stars rotating at lower frequencies ($346$~Hz and $716$~Hz), $I$ increases as the star heats up but decreases as it cools. Analyzing the variation of $M$ against $I$, we observe that all our results fall within the inferred value of PSR J0030$+$0451 reported in~\cite{silva2021PhRvL} for all baryonic composition, for PSR J0737$-$3039A only the curves with $346$~Hz satisfy the constraint in~\cite{bejger2005MNRAS}, which is a good result since this pulsar has a low rotation rate. On the other hand, only the mass-shedding limit curves match the constraints for PSR J0740$+$6620~\cite{li2022CQGra}, our estimate is that $I$ for a $2$~\msun~NS rotating at $346$~Hz is around $\rm \sim 3 \times 10^{45}~g cm^2$. We also compared our results with the universal relations between $I/(MR_e^2)$ and $M/R_e$ proposed by Breu \& Rezzolla~\cite{breu2016} and Steiner \textit{et al.}~\cite{steiner2015} and concluded that these relations must be shifted towards smaller values for hot and rapidly rotating NSs. 

Finally, we presented a summary of the properties of the PNSs with $\rm M_{max}$ rotating at the mass-shedding limit and conclude that the central temperature $T_c$ is the property most sensitive to the emergence of $\Delta$-resonances in the stellar core, with a decrease in $T_c$ when $\Delta$ baryons are added. In contrast, the ratio of the polar to equatorial radius $R_p / R_e$ is the least affected by the different EoS types. 

On the second part of this work, we are going to investigate the influence of rotation on the local properties of individual rotating PNSs along their evolutionary stages. In addition, we are going to analyze if the decrease in angular momentum due to neutrino loss can have a significative effect on the stages of evolution of the PNSs considered in our work. In this way, we will be able to investigate how the internal properties of the star -- such as particle fractions and temperature distribution, for example -- change with rotation and the presence of the $\Delta$-resonances.

\section*{Acknowledgements}
This work is part of the project INCT-FNA (Proc. No. 464898/2014-5) and is also supported by the National Council for Scientific and Technological Development (CNPq) under Grants Nos. 303490/2021-7 (D.P.M.) and 306834/2022-7  (T.F.). A.I. acknowledges financial support from the São Paulo Research Foundation (FAPESP), Grant No. 2023/09545-1. T.F. thanks the Brazilian Federal Agency for Support and Evaluation of Graduate Education (CAPES) (Finance Code 001) and FAPESP (Grants 2019/07767-1) for their financial support. L.C.N.S. would like to thank the Santa Catarina Research Foundation (FAPESC) for financial support under Grant No. 735/2024.

\bibliography{reference}

\begin{thebibliography}{102}%
\makeatletter
\providecommand \@ifxundefined [1]{%
 \@ifx{#1\undefined}
}%
\providecommand \@ifnum [1]{%
 \ifnum #1\expandafter \@firstoftwo
 \else \expandafter \@secondoftwo
 \fi
}%
\providecommand \@ifx [1]{%
 \ifx #1\expandafter \@firstoftwo
 \else \expandafter \@secondoftwo
 \fi
}%
\providecommand \natexlab [1]{#1}%
\providecommand \enquote  [1]{``#1''}%
\providecommand \bibnamefont  [1]{#1}%
\providecommand \bibfnamefont [1]{#1}%
\providecommand \citenamefont [1]{#1}%
\providecommand \href@noop [0]{\@secondoftwo}%
\providecommand \href [0]{\begingroup \@sanitize@url \@href}%
\providecommand \@href[1]{\@@startlink{#1}\@@href}%
\providecommand \@@href[1]{\endgroup#1\@@endlink}%
\providecommand \@sanitize@url [0]{\catcode `\\12\catcode `\$12\catcode
  `\&12\catcode `\#12\catcode `\^12\catcode `\_12\catcode `\%12\relax}%
\providecommand \@@startlink[1]{}%
\providecommand \@@endlink[0]{}%
\providecommand \url  [0]{\begingroup\@sanitize@url \@url }%
\providecommand \@url [1]{\endgroup\@href {#1}{\urlprefix }}%
\providecommand \urlprefix  [0]{URL }%
\providecommand \Eprint [0]{\href }%
\providecommand \doibase [0]{https://doi.org/}%
\providecommand \selectlanguage [0]{\@gobble}%
\providecommand \bibinfo  [0]{\@secondoftwo}%
\providecommand \bibfield  [0]{\@secondoftwo}%
\providecommand \translation [1]{[#1]}%
\providecommand \BibitemOpen [0]{}%
\providecommand \bibitemStop [0]{}%
\providecommand \bibitemNoStop [0]{.\EOS\space}%
\providecommand \EOS [0]{\spacefactor3000\relax}%
\providecommand \BibitemShut  [1]{\csname bibitem#1\endcsname}%
\let\auto@bib@innerbib\@empty
\bibitem [{\citenamefont {{Pons}}\ \emph {et~al.}(1999)\citenamefont {{Pons}},
  \citenamefont {{Reddy}}, \citenamefont {{Prakash}}, \citenamefont
  {{Lattimer}},\ and\ \citenamefont {{Miralles}}}]{Pons:1998mm}%
  \BibitemOpen
  \bibfield  {author} {\bibinfo {author} {\bibfnamefont {J.~A.}\ \bibnamefont
  {{Pons}}}, \bibinfo {author} {\bibfnamefont {S.}~\bibnamefont {{Reddy}}},
  \bibinfo {author} {\bibfnamefont {M.}~\bibnamefont {{Prakash}}}, \bibinfo
  {author} {\bibfnamefont {J.~M.}\ \bibnamefont {{Lattimer}}},\ and\ \bibinfo
  {author} {\bibfnamefont {J.~A.}\ \bibnamefont {{Miralles}}},\ }\href
  {https://doi.org/10.1086/306889} {\bibfield  {journal} {\bibinfo  {journal}
  {\apj}\ }\textbf {\bibinfo {volume} {513}},\ \bibinfo {pages} {780} (\bibinfo
  {year} {1999})},\ \Eprint {https://arxiv.org/abs/astro-ph/9807040}
  {arXiv:astro-ph/9807040 [astro-ph]} \BibitemShut {NoStop}%
\bibitem [{\citenamefont {{Prakash}}\ \emph {et~al.}(1997)\citenamefont
  {{Prakash}}, \citenamefont {{Bombaci}}, \citenamefont {{Prakash}},
  \citenamefont {{Ellis}}, \citenamefont {{Lattimer}},\ and\ \citenamefont
  {{Knorren}}}]{Prakash:1996xs}%
  \BibitemOpen
  \bibfield  {author} {\bibinfo {author} {\bibfnamefont {M.}~\bibnamefont
  {{Prakash}}}, \bibinfo {author} {\bibfnamefont {I.}~\bibnamefont
  {{Bombaci}}}, \bibinfo {author} {\bibfnamefont {M.}~\bibnamefont
  {{Prakash}}}, \bibinfo {author} {\bibfnamefont {P.~J.}\ \bibnamefont
  {{Ellis}}}, \bibinfo {author} {\bibfnamefont {J.~M.}\ \bibnamefont
  {{Lattimer}}},\ and\ \bibinfo {author} {\bibfnamefont {R.}~\bibnamefont
  {{Knorren}}},\ }\href {https://doi.org/10.1016/S0370-1573(96)00023-3}
  {\bibfield  {journal} {\bibinfo  {journal} {\physrep}\ }\textbf {\bibinfo
  {volume} {280}},\ \bibinfo {pages} {1} (\bibinfo {year} {1997})},\ \Eprint
  {https://arxiv.org/abs/nucl-th/9603042} {arXiv:nucl-th/9603042 [nucl-th]}
  \BibitemShut {NoStop}%
\bibitem [{\citenamefont {{Burrows}}\ and\ \citenamefont
  {{Lattimer}}(1986)}]{burrows1986}%
  \BibitemOpen
  \bibfield  {author} {\bibinfo {author} {\bibfnamefont {A.}~\bibnamefont
  {{Burrows}}}\ and\ \bibinfo {author} {\bibfnamefont {J.~M.}\ \bibnamefont
  {{Lattimer}}},\ }\href {https://doi.org/10.1086/164405} {\bibfield  {journal}
  {\bibinfo  {journal} {\apj}\ }\textbf {\bibinfo {volume} {307}},\ \bibinfo
  {pages} {178} (\bibinfo {year} {1986})}\BibitemShut {NoStop}%
\bibitem [{\citenamefont {{Pons}}\ \emph
  {et~al.}(2001{\natexlab{a}})\citenamefont {{Pons}}, \citenamefont
  {{Steiner}}, \citenamefont {{Prakash}},\ and\ \citenamefont
  {{Lattimer}}}]{Pons:2001ar}%
  \BibitemOpen
  \bibfield  {author} {\bibinfo {author} {\bibfnamefont {J.~A.}\ \bibnamefont
  {{Pons}}}, \bibinfo {author} {\bibfnamefont {A.~W.}\ \bibnamefont
  {{Steiner}}}, \bibinfo {author} {\bibfnamefont {M.}~\bibnamefont
  {{Prakash}}},\ and\ \bibinfo {author} {\bibfnamefont {J.~M.}\ \bibnamefont
  {{Lattimer}}},\ }\href {https://doi.org/10.1103/PhysRevLett.86.5223}
  {\bibfield  {journal} {\bibinfo  {journal} {\prl}\ }\textbf {\bibinfo
  {volume} {86}},\ \bibinfo {pages} {5223} (\bibinfo {year}
  {2001}{\natexlab{a}})},\ \Eprint {https://arxiv.org/abs/astro-ph/0102015}
  {arXiv:astro-ph/0102015 [astro-ph]} \BibitemShut {NoStop}%
\bibitem [{\citenamefont {{Pons}}\ \emph
  {et~al.}(2001{\natexlab{b}})\citenamefont {{Pons}}, \citenamefont
  {{Miralles}}, \citenamefont {{Prakash}},\ and\ \citenamefont
  {{Lattimer}}}]{Pons:2000xf}%
  \BibitemOpen
  \bibfield  {author} {\bibinfo {author} {\bibfnamefont {J.~A.}\ \bibnamefont
  {{Pons}}}, \bibinfo {author} {\bibfnamefont {J.~A.}\ \bibnamefont
  {{Miralles}}}, \bibinfo {author} {\bibfnamefont {M.}~\bibnamefont
  {{Prakash}}},\ and\ \bibinfo {author} {\bibfnamefont {J.~M.}\ \bibnamefont
  {{Lattimer}}},\ }\href {https://doi.org/10.1086/320642} {\bibfield  {journal}
  {\bibinfo  {journal} {\apj}\ }\textbf {\bibinfo {volume} {553}},\ \bibinfo
  {pages} {382} (\bibinfo {year} {2001}{\natexlab{b}})},\ \Eprint
  {https://arxiv.org/abs/astro-ph/0008389} {arXiv:astro-ph/0008389 [astro-ph]}
  \BibitemShut {NoStop}%
\bibitem [{\citenamefont {{Keil}}\ and\ \citenamefont
  {{Janka}}(1995)}]{keil1995}%
  \BibitemOpen
  \bibfield  {author} {\bibinfo {author} {\bibfnamefont {W.}~\bibnamefont
  {{Keil}}}\ and\ \bibinfo {author} {\bibfnamefont {H.~T.}\ \bibnamefont
  {{Janka}}},\ }\href@noop {} {\bibfield  {journal} {\bibinfo  {journal}
  {A\&A}\ }\textbf {\bibinfo {volume} {296}},\ \bibinfo {pages} {145} (\bibinfo
  {year} {1995})}\BibitemShut {NoStop}%
\bibitem [{\citenamefont {{van Belle}}(2012)}]{vanbelle2012A&ARv}%
  \BibitemOpen
  \bibfield  {author} {\bibinfo {author} {\bibfnamefont {G.~T.}\ \bibnamefont
  {{van Belle}}},\ }\href {https://doi.org/10.1007/s00159-012-0051-2}
  {\bibfield  {journal} {\bibinfo  {journal} {\aapr}\ }\textbf {\bibinfo
  {volume} {20}},\ \bibinfo {eid} {51} (\bibinfo {year} {2012})},\ \Eprint
  {https://arxiv.org/abs/1204.2572} {arXiv:1204.2572 [astro-ph.SR]}
  \BibitemShut {NoStop}%
\bibitem [{\citenamefont {{Fukuda}}(1982)}]{fukuda1982PASP}%
  \BibitemOpen
  \bibfield  {author} {\bibinfo {author} {\bibfnamefont {I.}~\bibnamefont
  {{Fukuda}}},\ }\href {https://doi.org/10.1086/130977} {\bibfield  {journal}
  {\bibinfo  {journal} {\pasp}\ }\textbf {\bibinfo {volume} {94}},\ \bibinfo
  {pages} {271} (\bibinfo {year} {1982})}\BibitemShut {NoStop}%
\bibitem [{\citenamefont {{Langer}}(2012)}]{langer2012ARA&A}%
  \BibitemOpen
  \bibfield  {author} {\bibinfo {author} {\bibfnamefont {N.}~\bibnamefont
  {{Langer}}},\ }\href {https://doi.org/10.1146/annurev-astro-081811-125534}
  {\bibfield  {journal} {\bibinfo  {journal} {\araa}\ }\textbf {\bibinfo
  {volume} {50}},\ \bibinfo {pages} {107} (\bibinfo {year} {2012})},\ \Eprint
  {https://arxiv.org/abs/1206.5443} {arXiv:1206.5443 [astro-ph.SR]}
  \BibitemShut {NoStop}%
\bibitem [{\citenamefont {{Woosley}}\ and\ \citenamefont
  {{Bloom}}(2006)}]{woosley2006ARA&A}%
  \BibitemOpen
  \bibfield  {author} {\bibinfo {author} {\bibfnamefont {S.~E.}\ \bibnamefont
  {{Woosley}}}\ and\ \bibinfo {author} {\bibfnamefont {J.~S.}\ \bibnamefont
  {{Bloom}}},\ }\href {https://doi.org/10.1146/annurev.astro.43.072103.150558}
  {\bibfield  {journal} {\bibinfo  {journal} {\araa}\ }\textbf {\bibinfo
  {volume} {44}},\ \bibinfo {pages} {507} (\bibinfo {year} {2006})},\ \Eprint
  {https://arxiv.org/abs/astro-ph/0609142} {arXiv:astro-ph/0609142 [astro-ph]}
  \BibitemShut {NoStop}%
\bibitem [{\citenamefont {{Nakamura}}\ \emph {et~al.}(2014)\citenamefont
  {{Nakamura}}, \citenamefont {{Kuroda}}, \citenamefont {{Takiwaki}},\ and\
  \citenamefont {{Kotake}}}]{nakamura2014ApJ}%
  \BibitemOpen
  \bibfield  {author} {\bibinfo {author} {\bibfnamefont {K.}~\bibnamefont
  {{Nakamura}}}, \bibinfo {author} {\bibfnamefont {T.}~\bibnamefont
  {{Kuroda}}}, \bibinfo {author} {\bibfnamefont {T.}~\bibnamefont
  {{Takiwaki}}},\ and\ \bibinfo {author} {\bibfnamefont {K.}~\bibnamefont
  {{Kotake}}},\ }\href {https://doi.org/10.1088/0004-637X/793/1/45} {\bibfield
  {journal} {\bibinfo  {journal} {\apj}\ }\textbf {\bibinfo {volume} {793}},\
  \bibinfo {eid} {45} (\bibinfo {year} {2014})},\ \Eprint
  {https://arxiv.org/abs/1403.7290} {arXiv:1403.7290 [astro-ph.HE]}
  \BibitemShut {NoStop}%
\bibitem [{\citenamefont {{M{\"o}sta}}\ \emph {et~al.}(2014)\citenamefont
  {{M{\"o}sta}}, \citenamefont {{Richers}}, \citenamefont {{Ott}},
  \citenamefont {{Haas}}, \citenamefont {{Piro}}, \citenamefont {{Boydstun}},
  \citenamefont {{Abdikamalov}}, \citenamefont {{Reisswig}},\ and\
  \citenamefont {{Schnetter}}}]{mosta2014ApJ}%
  \BibitemOpen
  \bibfield  {author} {\bibinfo {author} {\bibfnamefont {P.}~\bibnamefont
  {{M{\"o}sta}}}, \bibinfo {author} {\bibfnamefont {S.}~\bibnamefont
  {{Richers}}}, \bibinfo {author} {\bibfnamefont {C.~D.}\ \bibnamefont
  {{Ott}}}, \bibinfo {author} {\bibfnamefont {R.}~\bibnamefont {{Haas}}},
  \bibinfo {author} {\bibfnamefont {A.~L.}\ \bibnamefont {{Piro}}}, \bibinfo
  {author} {\bibfnamefont {K.}~\bibnamefont {{Boydstun}}}, \bibinfo {author}
  {\bibfnamefont {E.}~\bibnamefont {{Abdikamalov}}}, \bibinfo {author}
  {\bibfnamefont {C.}~\bibnamefont {{Reisswig}}},\ and\ \bibinfo {author}
  {\bibfnamefont {E.}~\bibnamefont {{Schnetter}}},\ }\href
  {https://doi.org/10.1088/2041-8205/785/2/L29} {\bibfield  {journal} {\bibinfo
   {journal} {\apjl}\ }\textbf {\bibinfo {volume} {785}},\ \bibinfo {eid} {L29}
  (\bibinfo {year} {2014})},\ \Eprint {https://arxiv.org/abs/1403.1230}
  {arXiv:1403.1230 [astro-ph.HE]} \BibitemShut {NoStop}%
\bibitem [{\citenamefont {{Summa}}\ \emph {et~al.}(2018)\citenamefont
  {{Summa}}, \citenamefont {{Janka}}, \citenamefont {{Melson}},\ and\
  \citenamefont {{Marek}}}]{summa2018ApJ}%
  \BibitemOpen
  \bibfield  {author} {\bibinfo {author} {\bibfnamefont {A.}~\bibnamefont
  {{Summa}}}, \bibinfo {author} {\bibfnamefont {H.-T.}\ \bibnamefont
  {{Janka}}}, \bibinfo {author} {\bibfnamefont {T.}~\bibnamefont {{Melson}}},\
  and\ \bibinfo {author} {\bibfnamefont {A.}~\bibnamefont {{Marek}}},\ }\href
  {https://doi.org/10.3847/1538-4357/aa9ce8} {\bibfield  {journal} {\bibinfo
  {journal} {\apj}\ }\textbf {\bibinfo {volume} {852}},\ \bibinfo {eid} {28}
  (\bibinfo {year} {2018})},\ \Eprint {https://arxiv.org/abs/1708.04154}
  {arXiv:1708.04154 [astro-ph.HE]} \BibitemShut {NoStop}%
\bibitem [{\citenamefont {{Thompson}}\ \emph {et~al.}(2005)\citenamefont
  {{Thompson}}, \citenamefont {{Quataert}},\ and\ \citenamefont
  {{Burrows}}}]{thompson2005ApJ}%
  \BibitemOpen
  \bibfield  {author} {\bibinfo {author} {\bibfnamefont {T.~A.}\ \bibnamefont
  {{Thompson}}}, \bibinfo {author} {\bibfnamefont {E.}~\bibnamefont
  {{Quataert}}},\ and\ \bibinfo {author} {\bibfnamefont {A.}~\bibnamefont
  {{Burrows}}},\ }\href {https://doi.org/10.1086/427177} {\bibfield  {journal}
  {\bibinfo  {journal} {\apj}\ }\textbf {\bibinfo {volume} {620}},\ \bibinfo
  {pages} {861} (\bibinfo {year} {2005})},\ \Eprint
  {https://arxiv.org/abs/astro-ph/0403224} {arXiv:astro-ph/0403224 [astro-ph]}
  \BibitemShut {NoStop}%
\bibitem [{\citenamefont {{Heger}}\ \emph {et~al.}(2005)\citenamefont
  {{Heger}}, \citenamefont {{Woosley}},\ and\ \citenamefont
  {{Spruit}}}]{heger2005ApJ}%
  \BibitemOpen
  \bibfield  {author} {\bibinfo {author} {\bibfnamefont {A.}~\bibnamefont
  {{Heger}}}, \bibinfo {author} {\bibfnamefont {S.~E.}\ \bibnamefont
  {{Woosley}}},\ and\ \bibinfo {author} {\bibfnamefont {H.~C.}\ \bibnamefont
  {{Spruit}}},\ }\href {https://doi.org/10.1086/429868} {\bibfield  {journal}
  {\bibinfo  {journal} {\apj}\ }\textbf {\bibinfo {volume} {626}},\ \bibinfo
  {pages} {350} (\bibinfo {year} {2005})},\ \Eprint
  {https://arxiv.org/abs/astro-ph/0409422} {arXiv:astro-ph/0409422 [astro-ph]}
  \BibitemShut {NoStop}%
\bibitem [{\citenamefont {{Hirschi}}\ \emph {et~al.}(2004)\citenamefont
  {{Hirschi}}, \citenamefont {{Meynet}},\ and\ \citenamefont
  {{Maeder}}}]{hirschi2004A&A}%
  \BibitemOpen
  \bibfield  {author} {\bibinfo {author} {\bibfnamefont {R.}~\bibnamefont
  {{Hirschi}}}, \bibinfo {author} {\bibfnamefont {G.}~\bibnamefont
  {{Meynet}}},\ and\ \bibinfo {author} {\bibfnamefont {A.}~\bibnamefont
  {{Maeder}}},\ }\href {https://doi.org/10.1051/0004-6361:20041095} {\bibfield
  {journal} {\bibinfo  {journal} {A\&A}\ }\textbf {\bibinfo {volume} {425}},\
  \bibinfo {pages} {649} (\bibinfo {year} {2004})},\ \Eprint
  {https://arxiv.org/abs/astro-ph/0406552} {arXiv:astro-ph/0406552 [astro-ph]}
  \BibitemShut {NoStop}%
\bibitem [{\citenamefont {{Janka}}(2004)}]{janka2004}%
  \BibitemOpen
  \bibfield  {author} {\bibinfo {author} {\bibfnamefont {H.~T.}\ \bibnamefont
  {{Janka}}},\ }in\ \href {https://doi.org/10.48550/arXiv.astro-ph/0402200}
  {\emph {\bibinfo {booktitle} {Young Neutron Stars and Their Environments}}},\
  \bibinfo {series} {IAU Symposium}, Vol.\ \bibinfo {volume} {218},\ \bibinfo
  {editor} {edited by\ \bibinfo {editor} {\bibfnamefont {F.}~\bibnamefont
  {{Camilo}}}\ and\ \bibinfo {editor} {\bibfnamefont {B.~M.}\ \bibnamefont
  {{Gaensler}}}}\ (\bibinfo {year} {2004})\ p.~\bibinfo {pages} {3},\ \Eprint
  {https://arxiv.org/abs/astro-ph/0402200} {arXiv:astro-ph/0402200 [astro-ph]}
  \BibitemShut {NoStop}%
\bibitem [{\citenamefont {{Martinon}}\ \emph {et~al.}(2014)\citenamefont
  {{Martinon}}, \citenamefont {{Maselli}}, \citenamefont {{Gualtieri}},\ and\
  \citenamefont {{Ferrari}}}]{martinon2014PhRvD}%
  \BibitemOpen
  \bibfield  {author} {\bibinfo {author} {\bibfnamefont {G.}~\bibnamefont
  {{Martinon}}}, \bibinfo {author} {\bibfnamefont {A.}~\bibnamefont
  {{Maselli}}}, \bibinfo {author} {\bibfnamefont {L.}~\bibnamefont
  {{Gualtieri}}},\ and\ \bibinfo {author} {\bibfnamefont {V.}~\bibnamefont
  {{Ferrari}}},\ }\href {https://doi.org/10.1103/PhysRevD.90.064026} {\bibfield
   {journal} {\bibinfo  {journal} {\prd}\ }\textbf {\bibinfo {volume} {90}},\
  \bibinfo {eid} {064026} (\bibinfo {year} {2014})},\ \Eprint
  {https://arxiv.org/abs/1406.7661} {arXiv:1406.7661 [gr-qc]} \BibitemShut
  {NoStop}%
\bibitem [{\citenamefont {{Andersson}}(2003)}]{andersson2003CQGra}%
  \BibitemOpen
  \bibfield  {author} {\bibinfo {author} {\bibfnamefont {N.}~\bibnamefont
  {{Andersson}}},\ }\href {https://doi.org/10.1088/0264-9381/20/7/201}
  {\bibfield  {journal} {\bibinfo  {journal} {Classical and Quantum Gravity}\
  }\textbf {\bibinfo {volume} {20}},\ \bibinfo {pages} {R105} (\bibinfo {year}
  {2003})},\ \Eprint {https://arxiv.org/abs/astro-ph/0211057}
  {arXiv:astro-ph/0211057 [astro-ph]} \BibitemShut {NoStop}%
\bibitem [{\citenamefont {{Thompson}}\ \emph {et~al.}(2004)\citenamefont
  {{Thompson}}, \citenamefont {{Chang}},\ and\ \citenamefont
  {{Quataert}}}]{thompson2004ApJ}%
  \BibitemOpen
  \bibfield  {author} {\bibinfo {author} {\bibfnamefont {T.~A.}\ \bibnamefont
  {{Thompson}}}, \bibinfo {author} {\bibfnamefont {P.}~\bibnamefont
  {{Chang}}},\ and\ \bibinfo {author} {\bibfnamefont {E.}~\bibnamefont
  {{Quataert}}},\ }\href {https://doi.org/10.1086/421969} {\bibfield  {journal}
  {\bibinfo  {journal} {\apj}\ }\textbf {\bibinfo {volume} {611}},\ \bibinfo
  {pages} {380} (\bibinfo {year} {2004})},\ \Eprint
  {https://arxiv.org/abs/astro-ph/0401555} {arXiv:astro-ph/0401555 [astro-ph]}
  \BibitemShut {NoStop}%
\bibitem [{\citenamefont {{Ott}}\ \emph {et~al.}(2006)\citenamefont {{Ott}},
  \citenamefont {{Burrows}}, \citenamefont {{Thompson}}, \citenamefont
  {{Livne}},\ and\ \citenamefont {{Walder}}}]{ott2006ApJS}%
  \BibitemOpen
  \bibfield  {author} {\bibinfo {author} {\bibfnamefont {C.~D.}\ \bibnamefont
  {{Ott}}}, \bibinfo {author} {\bibfnamefont {A.}~\bibnamefont {{Burrows}}},
  \bibinfo {author} {\bibfnamefont {T.~A.}\ \bibnamefont {{Thompson}}},
  \bibinfo {author} {\bibfnamefont {E.}~\bibnamefont {{Livne}}},\ and\ \bibinfo
  {author} {\bibfnamefont {R.}~\bibnamefont {{Walder}}},\ }\href
  {https://doi.org/10.1086/500832} {\bibfield  {journal} {\bibinfo  {journal}
  {\apjs}\ }\textbf {\bibinfo {volume} {164}},\ \bibinfo {pages} {130}
  (\bibinfo {year} {2006})},\ \Eprint {https://arxiv.org/abs/astro-ph/0508462}
  {arXiv:astro-ph/0508462 [astro-ph]} \BibitemShut {NoStop}%
\bibitem [{\citenamefont {{Marek}}\ and\ \citenamefont
  {{Janka}}(2009)}]{marek2009ApJ}%
  \BibitemOpen
  \bibfield  {author} {\bibinfo {author} {\bibfnamefont {A.}~\bibnamefont
  {{Marek}}}\ and\ \bibinfo {author} {\bibfnamefont {H.~T.}\ \bibnamefont
  {{Janka}}},\ }\href {https://doi.org/10.1088/0004-637X/694/1/664} {\bibfield
  {journal} {\bibinfo  {journal} {\apj}\ }\textbf {\bibinfo {volume} {694}},\
  \bibinfo {pages} {664} (\bibinfo {year} {2009})},\ \Eprint
  {https://arxiv.org/abs/0708.3372} {arXiv:0708.3372 [astro-ph]} \BibitemShut
  {NoStop}%
\bibitem [{\citenamefont {{Takiwaki}}\ \emph {et~al.}(2014)\citenamefont
  {{Takiwaki}}, \citenamefont {{Kotake}},\ and\ \citenamefont
  {{Suwa}}}]{takiwaki2014ApJ}%
  \BibitemOpen
  \bibfield  {author} {\bibinfo {author} {\bibfnamefont {T.}~\bibnamefont
  {{Takiwaki}}}, \bibinfo {author} {\bibfnamefont {K.}~\bibnamefont
  {{Kotake}}},\ and\ \bibinfo {author} {\bibfnamefont {Y.}~\bibnamefont
  {{Suwa}}},\ }\href {https://doi.org/10.1088/0004-637X/786/2/83} {\bibfield
  {journal} {\bibinfo  {journal} {\apj}\ }\textbf {\bibinfo {volume} {786}},\
  \bibinfo {eid} {83} (\bibinfo {year} {2014})},\ \Eprint
  {https://arxiv.org/abs/1308.5755} {arXiv:1308.5755 [astro-ph.SR]}
  \BibitemShut {NoStop}%
\bibitem [{\citenamefont {{Franzon}}\ \emph {et~al.}(2016)\citenamefont
  {{Franzon}}, \citenamefont {{Dexheimer}},\ and\ \citenamefont
  {{Schramm}}}]{franzon2016}%
  \BibitemOpen
  \bibfield  {author} {\bibinfo {author} {\bibfnamefont {B.}~\bibnamefont
  {{Franzon}}}, \bibinfo {author} {\bibfnamefont {V.}~\bibnamefont
  {{Dexheimer}}},\ and\ \bibinfo {author} {\bibfnamefont {S.}~\bibnamefont
  {{Schramm}}},\ }\href {https://doi.org/10.1103/PhysRevD.94.044018} {\bibfield
   {journal} {\bibinfo  {journal} {\prd}\ }\textbf {\bibinfo {volume} {94}},\
  \bibinfo {eid} {044018} (\bibinfo {year} {2016})},\ \Eprint
  {https://arxiv.org/abs/1606.04843} {arXiv:1606.04843 [astro-ph.HE]}
  \BibitemShut {NoStop}%
\bibitem [{\citenamefont {{Dexheimer}}\ and\ \citenamefont
  {{Schramm}}(2008)}]{Dexheimer:2008ax}%
  \BibitemOpen
  \bibfield  {author} {\bibinfo {author} {\bibfnamefont {V.}~\bibnamefont
  {{Dexheimer}}}\ and\ \bibinfo {author} {\bibfnamefont {S.}~\bibnamefont
  {{Schramm}}},\ }\href {https://doi.org/10.1086/589735} {\bibfield  {journal}
  {\bibinfo  {journal} {\apj}\ }\textbf {\bibinfo {volume} {683}},\ \bibinfo
  {pages} {943} (\bibinfo {year} {2008})},\ \Eprint
  {https://arxiv.org/abs/0802.1999} {arXiv:0802.1999 [astro-ph]} \BibitemShut
  {NoStop}%
\bibitem [{\citenamefont {{Issifu}}\ \emph {et~al.}(2023)\citenamefont
  {{Issifu}}, \citenamefont {{Marquez}}, \citenamefont {{Pelicer}},\ and\
  \citenamefont {{Menezes}}}]{Issifu:2023qyi}%
  \BibitemOpen
  \bibfield  {author} {\bibinfo {author} {\bibfnamefont {A.}~\bibnamefont
  {{Issifu}}}, \bibinfo {author} {\bibfnamefont {K.~D.}\ \bibnamefont
  {{Marquez}}}, \bibinfo {author} {\bibfnamefont {M.~R.}\ \bibnamefont
  {{Pelicer}}},\ and\ \bibinfo {author} {\bibfnamefont {D.~P.}\ \bibnamefont
  {{Menezes}}},\ }\href {https://doi.org/10.1093/mnras/stad1198} {\bibfield
  {journal} {\bibinfo  {journal} {\mnras}\ }\textbf {\bibinfo {volume} {522}},\
  \bibinfo {pages} {3263} (\bibinfo {year} {2023})},\ \Eprint
  {https://arxiv.org/abs/2302.04364} {arXiv:2302.04364 [nucl-th]} \BibitemShut
  {NoStop}%
\bibitem [{\citenamefont {{Malfatti}}\ \emph {et~al.}(2019)\citenamefont
  {{Malfatti}}, \citenamefont {{Orsaria}}, \citenamefont {{Contrera}},
  \citenamefont {{Weber}},\ and\ \citenamefont
  {{Ranea-Sandoval}}}]{Malfatti:2019tpg}%
  \BibitemOpen
  \bibfield  {author} {\bibinfo {author} {\bibfnamefont {G.}~\bibnamefont
  {{Malfatti}}}, \bibinfo {author} {\bibfnamefont {M.~G.}\ \bibnamefont
  {{Orsaria}}}, \bibinfo {author} {\bibfnamefont {G.~A.}\ \bibnamefont
  {{Contrera}}}, \bibinfo {author} {\bibfnamefont {F.}~\bibnamefont
  {{Weber}}},\ and\ \bibinfo {author} {\bibfnamefont {I.~F.}\ \bibnamefont
  {{Ranea-Sandoval}}},\ }\href {https://doi.org/10.1103/PhysRevC.100.015803}
  {\bibfield  {journal} {\bibinfo  {journal} {\prc}\ }\textbf {\bibinfo
  {volume} {100}},\ \bibinfo {eid} {015803} (\bibinfo {year} {2019})},\ \Eprint
  {https://arxiv.org/abs/1907.06597} {arXiv:1907.06597 [nucl-th]} \BibitemShut
  {NoStop}%
\bibitem [{\citenamefont {{Raduta}}\ \emph {et~al.}(2020)\citenamefont
  {{Raduta}}, \citenamefont {{Oertel}},\ and\ \citenamefont
  {{Sedrakian}}}]{Raduta:2020fdn}%
  \BibitemOpen
  \bibfield  {author} {\bibinfo {author} {\bibfnamefont {A.~R.}\ \bibnamefont
  {{Raduta}}}, \bibinfo {author} {\bibfnamefont {M.}~\bibnamefont {{Oertel}}},\
  and\ \bibinfo {author} {\bibfnamefont {A.}~\bibnamefont {{Sedrakian}}},\
  }\href {https://doi.org/10.1093/mnras/staa2491} {\bibfield  {journal}
  {\bibinfo  {journal} {\mnras}\ }\textbf {\bibinfo {volume} {499}},\ \bibinfo
  {pages} {914} (\bibinfo {year} {2020})},\ \Eprint
  {https://arxiv.org/abs/2008.00213} {arXiv:2008.00213 [nucl-th]} \BibitemShut
  {NoStop}%
\bibitem [{\citenamefont {{Ghosh}}\ \emph {et~al.}(2024)\citenamefont
  {{Ghosh}}, \citenamefont {{Shaikh}}, \citenamefont {{Kalita}}, \citenamefont
  {{Routaray}}, \citenamefont {{Kumar}},\ and\ \citenamefont
  {{Agrawal}}}]{Ghosh:2023tbn}%
  \BibitemOpen
  \bibfield  {author} {\bibinfo {author} {\bibfnamefont {S.}~\bibnamefont
  {{Ghosh}}}, \bibinfo {author} {\bibfnamefont {S.}~\bibnamefont {{Shaikh}}},
  \bibinfo {author} {\bibfnamefont {P.~J.}\ \bibnamefont {{Kalita}}}, \bibinfo
  {author} {\bibfnamefont {P.}~\bibnamefont {{Routaray}}}, \bibinfo {author}
  {\bibfnamefont {B.}~\bibnamefont {{Kumar}}},\ and\ \bibinfo {author}
  {\bibfnamefont {B.~K.}\ \bibnamefont {{Agrawal}}},\ }\href
  {https://doi.org/10.1016/j.nuclphysb.2024.116697} {\bibfield  {journal}
  {\bibinfo  {journal} {Nuclear Physics B}\ }\textbf {\bibinfo {volume}
  {1008}},\ \bibinfo {eid} {116697} (\bibinfo {year} {2024})}\BibitemShut
  {NoStop}%
\bibitem [{\citenamefont {{Villain}}\ \emph {et~al.}(2004)\citenamefont
  {{Villain}}, \citenamefont {{Pons}}, \citenamefont {{Cerd{\'a}-Dur{\'a}n}},\
  and\ \citenamefont {{Gourgoulhon}}}]{Villain:2003ey}%
  \BibitemOpen
  \bibfield  {author} {\bibinfo {author} {\bibfnamefont {L.}~\bibnamefont
  {{Villain}}}, \bibinfo {author} {\bibfnamefont {J.~A.}\ \bibnamefont
  {{Pons}}}, \bibinfo {author} {\bibfnamefont {P.}~\bibnamefont
  {{Cerd{\'a}-Dur{\'a}n}}},\ and\ \bibinfo {author} {\bibfnamefont
  {E.}~\bibnamefont {{Gourgoulhon}}},\ }\href
  {https://doi.org/10.1051/0004-6361:20035619} {\bibfield  {journal} {\bibinfo
  {journal} {A\&A}\ }\textbf {\bibinfo {volume} {418}},\ \bibinfo {pages} {283}
  (\bibinfo {year} {2004})},\ \Eprint {https://arxiv.org/abs/astro-ph/0310875}
  {arXiv:astro-ph/0310875 [astro-ph]} \BibitemShut {NoStop}%
\bibitem [{\citenamefont {{Riley}}\ \emph {et~al.}(2019)\citenamefont
  {{Riley}}, \citenamefont {{Watts}}, \citenamefont {{Bogdanov}} \emph
  {et~al.}}]{riley2019}%
  \BibitemOpen
  \bibfield  {author} {\bibinfo {author} {\bibfnamefont {T.~E.}\ \bibnamefont
  {{Riley}}}, \bibinfo {author} {\bibfnamefont {A.~L.}\ \bibnamefont
  {{Watts}}}, \bibinfo {author} {\bibfnamefont {S.}~\bibnamefont {{Bogdanov}}},
  \emph {et~al.},\ }\href {https://doi.org/10.3847/2041-8213/ab481c} {\bibfield
   {journal} {\bibinfo  {journal} {\apjl}\ }\textbf {\bibinfo {volume} {887}},\
  \bibinfo {eid} {L21} (\bibinfo {year} {2019})},\ \Eprint
  {https://arxiv.org/abs/1912.05702} {arXiv:1912.05702 [astro-ph.HE]}
  \BibitemShut {NoStop}%
\bibitem [{\citenamefont {{Miller}}\ \emph {et~al.}(2019)\citenamefont
  {{Miller}}, \citenamefont {{Lamb}}, \citenamefont {{Dittmann}} \emph
  {et~al.}}]{Miller:2019cac}%
  \BibitemOpen
  \bibfield  {author} {\bibinfo {author} {\bibfnamefont {M.~C.}\ \bibnamefont
  {{Miller}}}, \bibinfo {author} {\bibfnamefont {F.~K.}\ \bibnamefont
  {{Lamb}}}, \bibinfo {author} {\bibfnamefont {A.~J.}\ \bibnamefont
  {{Dittmann}}}, \emph {et~al.},\ }\href
  {https://doi.org/10.3847/2041-8213/ab50c5} {\bibfield  {journal} {\bibinfo
  {journal} {\apjl}\ }\textbf {\bibinfo {volume} {887}},\ \bibinfo {eid} {L24}
  (\bibinfo {year} {2019})},\ \Eprint {https://arxiv.org/abs/1912.05705}
  {arXiv:1912.05705 [astro-ph.HE]} \BibitemShut {NoStop}%
\bibitem [{\citenamefont {{Riley}}\ \emph {et~al.}(2021)\citenamefont
  {{Riley}}, \citenamefont {{Watts}}, \citenamefont {{Ray}} \emph
  {et~al.}}]{riley2021}%
  \BibitemOpen
  \bibfield  {author} {\bibinfo {author} {\bibfnamefont {T.~E.}\ \bibnamefont
  {{Riley}}}, \bibinfo {author} {\bibfnamefont {A.~L.}\ \bibnamefont
  {{Watts}}}, \bibinfo {author} {\bibfnamefont {P.~S.}\ \bibnamefont {{Ray}}},
  \emph {et~al.},\ }\href {https://doi.org/10.3847/2041-8213/ac0a81} {\bibfield
   {journal} {\bibinfo  {journal} {\apjl}\ }\textbf {\bibinfo {volume} {918}},\
  \bibinfo {eid} {L27} (\bibinfo {year} {2021})},\ \Eprint
  {https://arxiv.org/abs/2105.06980} {arXiv:2105.06980 [astro-ph.HE]}
  \BibitemShut {NoStop}%
\bibitem [{\citenamefont {{Miller}}\ \emph {et~al.}(2021)\citenamefont
  {{Miller}}, \citenamefont {{Lamb}}, \citenamefont {{Dittmann}} \emph
  {et~al.}}]{miller2021}%
  \BibitemOpen
  \bibfield  {author} {\bibinfo {author} {\bibfnamefont {M.~C.}\ \bibnamefont
  {{Miller}}}, \bibinfo {author} {\bibfnamefont {F.~K.}\ \bibnamefont
  {{Lamb}}}, \bibinfo {author} {\bibfnamefont {A.~J.}\ \bibnamefont
  {{Dittmann}}}, \emph {et~al.},\ }\href
  {https://doi.org/10.3847/2041-8213/ac089b} {\bibfield  {journal} {\bibinfo
  {journal} {\apjl}\ }\textbf {\bibinfo {volume} {918}},\ \bibinfo {eid} {L28}
  (\bibinfo {year} {2021})},\ \Eprint {https://arxiv.org/abs/2105.06979}
  {arXiv:2105.06979 [astro-ph.HE]} \BibitemShut {NoStop}%
\bibitem [{\citenamefont {{Choudhury}}\ \emph {et~al.}(2024)\citenamefont
  {{Choudhury}}, \citenamefont {{Salmi}}, \citenamefont {{Vinciguerra}},
  \citenamefont {{Riley}} \emph {et~al.}}]{choudhury2024ApJ}%
  \BibitemOpen
  \bibfield  {author} {\bibinfo {author} {\bibfnamefont {D.}~\bibnamefont
  {{Choudhury}}}, \bibinfo {author} {\bibfnamefont {T.}~\bibnamefont
  {{Salmi}}}, \bibinfo {author} {\bibfnamefont {S.}~\bibnamefont
  {{Vinciguerra}}}, \bibinfo {author} {\bibfnamefont {T.~E.}\ \bibnamefont
  {{Riley}}}, \emph {et~al.},\ }\href
  {https://doi.org/10.3847/2041-8213/ad5a6f} {\bibfield  {journal} {\bibinfo
  {journal} {\apjl}\ }\textbf {\bibinfo {volume} {971}},\ \bibinfo {eid} {L20}
  (\bibinfo {year} {2024})},\ \Eprint {https://arxiv.org/abs/2407.06789}
  {arXiv:2407.06789 [astro-ph.HE]} \BibitemShut {NoStop}%
\bibitem [{\citenamefont {{Salmi}}\ \emph {et~al.}(2024)\citenamefont
  {{Salmi}}, \citenamefont {{Deneva}}, \citenamefont {{Ray}}, \citenamefont
  {{Watts}} \emph {et~al.}}]{salmi2024ApJ}%
  \BibitemOpen
  \bibfield  {author} {\bibinfo {author} {\bibfnamefont {T.}~\bibnamefont
  {{Salmi}}}, \bibinfo {author} {\bibfnamefont {J.~S.}\ \bibnamefont
  {{Deneva}}}, \bibinfo {author} {\bibfnamefont {P.~S.}\ \bibnamefont {{Ray}}},
  \bibinfo {author} {\bibfnamefont {A.~L.}\ \bibnamefont {{Watts}}}, \emph
  {et~al.},\ }\href {https://doi.org/10.3847/1538-4357/ad81d2} {\bibfield
  {journal} {\bibinfo  {journal} {\apj}\ }\textbf {\bibinfo {volume} {976}},\
  \bibinfo {eid} {58} (\bibinfo {year} {2024})},\ \Eprint
  {https://arxiv.org/abs/2409.14923} {arXiv:2409.14923 [astro-ph.HE]}
  \BibitemShut {NoStop}%
\bibitem [{\citenamefont {{Abbott}}\ \emph
  {et~al.}(2017{\natexlab{a}})\citenamefont {{Abbott}} \emph
  {et~al.}}]{LIGOScientific:2017vwq}%
  \BibitemOpen
  \bibfield  {author} {\bibinfo {author} {\bibfnamefont {B.~P.}\ \bibnamefont
  {{Abbott}}} \emph {et~al.},\ }\href
  {https://doi.org/10.1103/PhysRevLett.119.161101} {\bibfield  {journal}
  {\bibinfo  {journal} {\prl}\ }\textbf {\bibinfo {volume} {119}},\ \bibinfo
  {eid} {161101} (\bibinfo {year} {2017}{\natexlab{a}})},\ \Eprint
  {https://arxiv.org/abs/1710.05832} {arXiv:1710.05832 [gr-qc]} \BibitemShut
  {NoStop}%
\bibitem [{\citenamefont {{Abbott}}\ \emph
  {et~al.}(2017{\natexlab{b}})\citenamefont {{Abbott}}, \citenamefont
  {{Abbott}}, \citenamefont {{Abbott}}, \citenamefont {{Acernese}} \emph
  {et~al.}}]{2017ApJ...848L..12A}%
  \BibitemOpen
  \bibfield  {author} {\bibinfo {author} {\bibfnamefont {B.~P.}\ \bibnamefont
  {{Abbott}}}, \bibinfo {author} {\bibfnamefont {R.}~\bibnamefont {{Abbott}}},
  \bibinfo {author} {\bibfnamefont {T.~D.}\ \bibnamefont {{Abbott}}}, \bibinfo
  {author} {\bibfnamefont {F.}~\bibnamefont {{Acernese}}}, \emph {et~al.},\
  }\href {https://doi.org/10.3847/2041-8213/aa91c9} {\bibfield  {journal}
  {\bibinfo  {journal} {\apjl}\ }\textbf {\bibinfo {volume} {848}},\ \bibinfo
  {eid} {L12} (\bibinfo {year} {2017}{\natexlab{b}})},\ \Eprint
  {https://arxiv.org/abs/1710.05833} {arXiv:1710.05833 [astro-ph.HE]}
  \BibitemShut {NoStop}%
\bibitem [{\citenamefont {{Abbott}}\ \emph
  {et~al.}(2017{\natexlab{c}})\citenamefont {{Abbott}}, \citenamefont
  {{Abbott}}, \citenamefont {{Abbott}}, \citenamefont {{Acernese}} \emph
  {et~al.}}]{2017ApJ...848L..13A}%
  \BibitemOpen
  \bibfield  {author} {\bibinfo {author} {\bibfnamefont {B.~P.}\ \bibnamefont
  {{Abbott}}}, \bibinfo {author} {\bibfnamefont {R.}~\bibnamefont {{Abbott}}},
  \bibinfo {author} {\bibfnamefont {T.~D.}\ \bibnamefont {{Abbott}}}, \bibinfo
  {author} {\bibfnamefont {F.}~\bibnamefont {{Acernese}}}, \emph {et~al.},\
  }\href {https://doi.org/10.3847/2041-8213/aa920c} {\bibfield  {journal}
  {\bibinfo  {journal} {\apjl}\ }\textbf {\bibinfo {volume} {848}},\ \bibinfo
  {eid} {L13} (\bibinfo {year} {2017}{\natexlab{c}})},\ \Eprint
  {https://arxiv.org/abs/1710.05834} {arXiv:1710.05834 [astro-ph.HE]}
  \BibitemShut {NoStop}%
\bibitem [{\citenamefont {{Abbott}}\ \emph {et~al.}(2018)\citenamefont
  {{Abbott}}, \citenamefont {{Abbott}}, \citenamefont {{Abbott}} \emph
  {et~al.}}]{gw172018}%
  \BibitemOpen
  \bibfield  {author} {\bibinfo {author} {\bibfnamefont {B.~P.}\ \bibnamefont
  {{Abbott}}}, \bibinfo {author} {\bibfnamefont {R.}~\bibnamefont {{Abbott}}},
  \bibinfo {author} {\bibfnamefont {T.~D.}\ \bibnamefont {{Abbott}}}, \emph
  {et~al.},\ }\href {https://doi.org/10.1103/PhysRevLett.121.161101} {\bibfield
   {journal} {\bibinfo  {journal} {\prl}\ }\textbf {\bibinfo {volume} {121}},\
  \bibinfo {eid} {161101} (\bibinfo {year} {2018})},\ \Eprint
  {https://arxiv.org/abs/1805.11581} {arXiv:1805.11581 [gr-qc]} \BibitemShut
  {NoStop}%
\bibitem [{\citenamefont {Issifu}\ \emph {et~al.}(2025)\citenamefont {Issifu},
  \citenamefont {Menezes}, \citenamefont {Rezaei},\ and\ \citenamefont
  {Frederico}}]{Issifu_2025}%
  \BibitemOpen
  \bibfield  {author} {\bibinfo {author} {\bibfnamefont {A.}~\bibnamefont
  {Issifu}}, \bibinfo {author} {\bibfnamefont {D.~P.}\ \bibnamefont {Menezes}},
  \bibinfo {author} {\bibfnamefont {Z.}~\bibnamefont {Rezaei}},\ and\ \bibinfo
  {author} {\bibfnamefont {T.}~\bibnamefont {Frederico}},\ }\href
  {https://doi.org/10.1088/1475-7516/2025/01/024} {\bibfield  {journal}
  {\bibinfo  {journal} {Journal of Cosmology and Astroparticle Physics}\
  }\textbf {\bibinfo {volume} {2025}}\bibinfo  {number} { (01)},\ \bibinfo
  {pages} {024}}\BibitemShut {NoStop}%
\bibitem [{\citenamefont {{Issifu}}\ \emph
  {et~al.}(2024{\natexlab{a}})\citenamefont {{Issifu}}, \citenamefont {{da
  Silva}},\ and\ \citenamefont {{Menezes}}}]{Issifu:2023ovi}%
  \BibitemOpen
\bibfield  {number} {  }\bibfield  {author} {\bibinfo {author} {\bibfnamefont
  {A.}~\bibnamefont {{Issifu}}}, \bibinfo {author} {\bibfnamefont {F.~M.}\
  \bibnamefont {{da Silva}}},\ and\ \bibinfo {author} {\bibfnamefont {D.~P.}\
  \bibnamefont {{Menezes}}},\ }\href
  {https://doi.org/10.1140/epjc/s10052-024-12828-0} {\bibfield  {journal}
  {\bibinfo  {journal} {European Physical Journal C}\ }\textbf {\bibinfo
  {volume} {84}},\ \bibinfo {eid} {463} (\bibinfo {year}
  {2024}{\natexlab{a}})},\ \Eprint {https://arxiv.org/abs/2311.12511}
  {arXiv:2311.12511 [nucl-th]} \BibitemShut {NoStop}%
\bibitem [{\citenamefont {Goussard}\ \emph {et~al.}(1997)\citenamefont
  {Goussard}, \citenamefont {Haensel},\ and\ \citenamefont
  {Zdunik}}]{Goussard:1996dp}%
  \BibitemOpen
  \bibfield  {author} {\bibinfo {author} {\bibfnamefont {J.-O.}\ \bibnamefont
  {Goussard}}, \bibinfo {author} {\bibfnamefont {P.}~\bibnamefont {Haensel}},\
  and\ \bibinfo {author} {\bibfnamefont {J.~L.}\ \bibnamefont {Zdunik}},\
  }\href@noop {} {\bibfield  {journal} {\bibinfo  {journal} {Astron.
  Astrophys.}\ }\textbf {\bibinfo {volume} {321}},\ \bibinfo {pages} {822}
  (\bibinfo {year} {1997})},\ \Eprint {https://arxiv.org/abs/astro-ph/9610265}
  {arXiv:astro-ph/9610265} \BibitemShut {NoStop}%
\bibitem [{\citenamefont {{Cook}}\ \emph {et~al.}(1992)\citenamefont {{Cook}},
  \citenamefont {{Shapiro}},\ and\ \citenamefont {{Teukolsky}}}]{cook1992ApJ}%
  \BibitemOpen
  \bibfield  {author} {\bibinfo {author} {\bibfnamefont {G.~B.}\ \bibnamefont
  {{Cook}}}, \bibinfo {author} {\bibfnamefont {S.~L.}\ \bibnamefont
  {{Shapiro}}},\ and\ \bibinfo {author} {\bibfnamefont {S.~A.}\ \bibnamefont
  {{Teukolsky}}},\ }\href {https://doi.org/10.1086/171849} {\bibfield
  {journal} {\bibinfo  {journal} {\apj}\ }\textbf {\bibinfo {volume} {398}},\
  \bibinfo {pages} {203} (\bibinfo {year} {1992})}\BibitemShut {NoStop}%
\bibitem [{\citenamefont {Bozzola}\ \emph {et~al.}(2018)\citenamefont
  {Bozzola}, \citenamefont {Stergioulas},\ and\ \citenamefont
  {Bauswein}}]{Bozzola:2017qbu}%
  \BibitemOpen
  \bibfield  {author} {\bibinfo {author} {\bibfnamefont {G.}~\bibnamefont
  {Bozzola}}, \bibinfo {author} {\bibfnamefont {N.}~\bibnamefont
  {Stergioulas}},\ and\ \bibinfo {author} {\bibfnamefont {A.}~\bibnamefont
  {Bauswein}},\ }\href {https://doi.org/10.1093/mnras/stx3002} {\bibfield
  {journal} {\bibinfo  {journal} {Mon. Not. Roy. Astron. Soc.}\ }\textbf
  {\bibinfo {volume} {474}},\ \bibinfo {pages} {3557} (\bibinfo {year}
  {2018})},\ \Eprint {https://arxiv.org/abs/1709.02787} {arXiv:1709.02787
  [gr-qc]} \BibitemShut {NoStop}%
\bibitem [{\citenamefont {Galeazzi}\ \emph {et~al.}(2012)\citenamefont
  {Galeazzi}, \citenamefont {Yoshida},\ and\ \citenamefont
  {Eriguchi}}]{Galeazzi:2011nn}%
  \BibitemOpen
  \bibfield  {author} {\bibinfo {author} {\bibfnamefont {F.}~\bibnamefont
  {Galeazzi}}, \bibinfo {author} {\bibfnamefont {S.}~\bibnamefont {Yoshida}},\
  and\ \bibinfo {author} {\bibfnamefont {Y.}~\bibnamefont {Eriguchi}},\ }\href
  {https://doi.org/10.1051/0004-6361/201016316} {\bibfield  {journal} {\bibinfo
   {journal} {Astron. Astrophys.}\ }\textbf {\bibinfo {volume} {541}},\
  \bibinfo {pages} {A156} (\bibinfo {year} {2012})},\ \Eprint
  {https://arxiv.org/abs/1101.2664} {arXiv:1101.2664 [astro-ph.SR]}
  \BibitemShut {NoStop}%
\bibitem [{\citenamefont {Paschalidis}\ and\ \citenamefont
  {Stergioulas}(2017)}]{Paschalidis:2016vmz}%
  \BibitemOpen
  \bibfield  {author} {\bibinfo {author} {\bibfnamefont {V.}~\bibnamefont
  {Paschalidis}}\ and\ \bibinfo {author} {\bibfnamefont {N.}~\bibnamefont
  {Stergioulas}},\ }\href {https://doi.org/10.1007/s41114-017-0008-x}
  {\bibfield  {journal} {\bibinfo  {journal} {Living Rev. Rel.}\ }\textbf
  {\bibinfo {volume} {20}},\ \bibinfo {pages} {7} (\bibinfo {year} {2017})},\
  \Eprint {https://arxiv.org/abs/1612.03050} {arXiv:1612.03050 [astro-ph.HE]}
  \BibitemShut {NoStop}%
\bibitem [{\citenamefont {de~Paoli}\ \emph {et~al.}(2013)\citenamefont
  {de~Paoli}, \citenamefont {Menezes}, \citenamefont {Castro},\ and\
  \citenamefont {Barros}}]{dePaoli:2012eq}%
  \BibitemOpen
  \bibfield  {author} {\bibinfo {author} {\bibfnamefont {M.~G.}\ \bibnamefont
  {de~Paoli}}, \bibinfo {author} {\bibfnamefont {D.~P.}\ \bibnamefont
  {Menezes}}, \bibinfo {author} {\bibfnamefont {L.~B.}\ \bibnamefont
  {Castro}},\ and\ \bibinfo {author} {\bibfnamefont {C.~C.}\ \bibnamefont
  {Barros}, \bibfnamefont {Jr}},\ }\href
  {https://doi.org/10.1088/0954-3899/40/5/055007} {\bibfield  {journal}
  {\bibinfo  {journal} {J. Phys. G}\ }\textbf {\bibinfo {volume} {40}},\
  \bibinfo {pages} {055007} (\bibinfo {year} {2013})},\ \Eprint
  {https://arxiv.org/abs/1207.4063} {arXiv:1207.4063 [math-ph]} \BibitemShut
  {NoStop}%
\bibitem [{\citenamefont {Bombaci}(2017)}]{Bombaci:2016xzl}%
  \BibitemOpen
  \bibfield  {author} {\bibinfo {author} {\bibfnamefont {I.}~\bibnamefont
  {Bombaci}},\ }\href {https://doi.org/10.7566/JPSCP.17.101002} {\bibfield
  {journal} {\bibinfo  {journal} {JPS Conf. Proc.}\ }\textbf {\bibinfo {volume}
  {17}},\ \bibinfo {pages} {101002} (\bibinfo {year} {2017})},\ \Eprint
  {https://arxiv.org/abs/1601.05339} {arXiv:1601.05339 [nucl-th]} \BibitemShut
  {NoStop}%
\bibitem [{\citenamefont {Menezes}(2021)}]{Menezes:2021jmw}%
  \BibitemOpen
  \bibfield  {author} {\bibinfo {author} {\bibfnamefont {D.~P.}\ \bibnamefont
  {Menezes}},\ }\href {https://doi.org/10.3390/universe7080267} {\bibfield
  {journal} {\bibinfo  {journal} {Universe}\ }\textbf {\bibinfo {volume} {7}},\
  \bibinfo {pages} {267} (\bibinfo {year} {2021})},\ \Eprint
  {https://arxiv.org/abs/2106.09515} {arXiv:2106.09515 [astro-ph.HE]}
  \BibitemShut {NoStop}%
\bibitem [{\citenamefont {Roca-Maza}\ \emph {et~al.}(2011)\citenamefont
  {Roca-Maza}, \citenamefont {Vinas}, \citenamefont {Centelles}, \citenamefont
  {Ring},\ and\ \citenamefont {Schuck}}]{Roca-Maza:2011alv}%
  \BibitemOpen
  \bibfield  {author} {\bibinfo {author} {\bibfnamefont {X.}~\bibnamefont
  {Roca-Maza}}, \bibinfo {author} {\bibfnamefont {X.}~\bibnamefont {Vinas}},
  \bibinfo {author} {\bibfnamefont {M.}~\bibnamefont {Centelles}}, \bibinfo
  {author} {\bibfnamefont {P.}~\bibnamefont {Ring}},\ and\ \bibinfo {author}
  {\bibfnamefont {P.}~\bibnamefont {Schuck}},\ }\href
  {https://doi.org/10.1103/PhysRevC.84.054309} {\bibfield  {journal} {\bibinfo
  {journal} {Phys. Rev. C}\ }\textbf {\bibinfo {volume} {84}},\ \bibinfo
  {pages} {054309} (\bibinfo {year} {2011})},\ \bibinfo {note} {[Erratum:
  Phys.Rev.C 93, 069905 (2016)]},\ \Eprint {https://arxiv.org/abs/1110.2311}
  {arXiv:1110.2311 [nucl-th]} \BibitemShut {NoStop}%
\bibitem [{\citenamefont {Reed}\ \emph {et~al.}(2021)\citenamefont {Reed},
  \citenamefont {Fattoyev}, \citenamefont {Horowitz},\ and\ \citenamefont
  {Piekarewicz}}]{Reed:2021nqk}%
  \BibitemOpen
  \bibfield  {author} {\bibinfo {author} {\bibfnamefont {B.~T.}\ \bibnamefont
  {Reed}}, \bibinfo {author} {\bibfnamefont {F.~J.}\ \bibnamefont {Fattoyev}},
  \bibinfo {author} {\bibfnamefont {C.~J.}\ \bibnamefont {Horowitz}},\ and\
  \bibinfo {author} {\bibfnamefont {J.}~\bibnamefont {Piekarewicz}},\ }\href
  {https://doi.org/10.1103/PhysRevLett.126.172503} {\bibfield  {journal}
  {\bibinfo  {journal} {Phys. Rev. Lett.}\ }\textbf {\bibinfo {volume} {126}},\
  \bibinfo {pages} {172503} (\bibinfo {year} {2021})},\ \Eprint
  {https://arxiv.org/abs/2101.03193} {arXiv:2101.03193 [nucl-th]} \BibitemShut
  {NoStop}%
\bibitem [{\citenamefont {Lattimer}(2023)}]{Lattimer:2023rpe}%
  \BibitemOpen
  \bibfield  {author} {\bibinfo {author} {\bibfnamefont {J.~M.}\ \bibnamefont
  {Lattimer}},\ }\href {https://doi.org/10.3390/particles6010003} {\bibfield
  {journal} {\bibinfo  {journal} {Particles}\ }\textbf {\bibinfo {volume}
  {6}},\ \bibinfo {pages} {30} (\bibinfo {year} {2023})},\ \Eprint
  {https://arxiv.org/abs/2301.03666} {arXiv:2301.03666 [nucl-th]} \BibitemShut
  {NoStop}%
\bibitem [{\citenamefont {Dutra}\ \emph {et~al.}(2014)\citenamefont {Dutra},
  \citenamefont {Louren\c{c}o}, \citenamefont {Avancini}, \citenamefont
  {Carlson}, \citenamefont {Delfino}, \citenamefont {Menezes}, \citenamefont
  {Provid\^encia}, \citenamefont {Typel},\ and\ \citenamefont
  {Stone}}]{Dutra:2014qga}%
  \BibitemOpen
  \bibfield  {author} {\bibinfo {author} {\bibfnamefont {M.}~\bibnamefont
  {Dutra}}, \bibinfo {author} {\bibfnamefont {O.}~\bibnamefont {Louren\c{c}o}},
  \bibinfo {author} {\bibfnamefont {S.~S.}\ \bibnamefont {Avancini}}, \bibinfo
  {author} {\bibfnamefont {B.~V.}\ \bibnamefont {Carlson}}, \bibinfo {author}
  {\bibfnamefont {A.}~\bibnamefont {Delfino}}, \bibinfo {author} {\bibfnamefont
  {D.~P.}\ \bibnamefont {Menezes}}, \bibinfo {author} {\bibfnamefont
  {C.}~\bibnamefont {Provid\^encia}}, \bibinfo {author} {\bibfnamefont
  {S.}~\bibnamefont {Typel}},\ and\ \bibinfo {author} {\bibfnamefont {J.~R.}\
  \bibnamefont {Stone}},\ }\href {https://doi.org/10.1103/PhysRevC.90.055203}
  {\bibfield  {journal} {\bibinfo  {journal} {Phys. Rev. C}\ }\textbf {\bibinfo
  {volume} {90}},\ \bibinfo {pages} {055203} (\bibinfo {year} {2014})},\
  \Eprint {https://arxiv.org/abs/1405.3633} {arXiv:1405.3633 [nucl-th]}
  \BibitemShut {NoStop}%
\bibitem [{\citenamefont {Glendenning}\ and\ \citenamefont
  {Moszkowski}(1991)}]{Glendenning:1991es}%
  \BibitemOpen
  \bibfield  {author} {\bibinfo {author} {\bibfnamefont {N.~K.}\ \bibnamefont
  {Glendenning}}\ and\ \bibinfo {author} {\bibfnamefont {S.~A.}\ \bibnamefont
  {Moszkowski}},\ }\href {https://doi.org/10.1103/PhysRevLett.67.2414}
  {\bibfield  {journal} {\bibinfo  {journal} {Phys. Rev. Lett.}\ }\textbf
  {\bibinfo {volume} {67}},\ \bibinfo {pages} {2414} (\bibinfo {year}
  {1991})}\BibitemShut {NoStop}%
\bibitem [{\citenamefont {Lopes}\ and\ \citenamefont
  {Menezes}(2021)}]{Lopes:2020rqn}%
  \BibitemOpen
  \bibfield  {author} {\bibinfo {author} {\bibfnamefont {L.~L.}\ \bibnamefont
  {Lopes}}\ and\ \bibinfo {author} {\bibfnamefont {D.~P.}\ \bibnamefont
  {Menezes}},\ }\href {https://doi.org/10.1016/j.nuclphysa.2021.122171}
  {\bibfield  {journal} {\bibinfo  {journal} {Nucl. Phys. A}\ }\textbf
  {\bibinfo {volume} {1009}},\ \bibinfo {pages} {122171} (\bibinfo {year}
  {2021})},\ \Eprint {https://arxiv.org/abs/2004.07909} {arXiv:2004.07909
  [astro-ph.HE]} \BibitemShut {NoStop}%
\bibitem [{\citenamefont {Pais}(1966)}]{Pais:1966eox}%
  \BibitemOpen
  \bibfield  {author} {\bibinfo {author} {\bibfnamefont {A.}~\bibnamefont
  {Pais}},\ }\href {https://doi.org/10.1103/revmodphys.38.215} {\bibfield
  {journal} {\bibinfo  {journal} {Rev. Mod. Phys.}\ }\textbf {\bibinfo {volume}
  {38}},\ \bibinfo {pages} {215} (\bibinfo {year} {1966})}\BibitemShut
  {NoStop}%
\bibitem [{\citenamefont {Weissenborn}\ \emph {et~al.}(2012)\citenamefont
  {Weissenborn}, \citenamefont {Chatterjee},\ and\ \citenamefont
  {Schaffner-Bielich}}]{Weissenborn:2011kb}%
  \BibitemOpen
  \bibfield  {author} {\bibinfo {author} {\bibfnamefont {S.}~\bibnamefont
  {Weissenborn}}, \bibinfo {author} {\bibfnamefont {D.}~\bibnamefont
  {Chatterjee}},\ and\ \bibinfo {author} {\bibfnamefont {J.}~\bibnamefont
  {Schaffner-Bielich}},\ }\href
  {https://doi.org/10.1016/j.nuclphysa.2012.02.012} {\bibfield  {journal}
  {\bibinfo  {journal} {Nucl. Phys. A}\ }\textbf {\bibinfo {volume} {881}},\
  \bibinfo {pages} {62} (\bibinfo {year} {2012})},\ \Eprint
  {https://arxiv.org/abs/1111.6049} {arXiv:1111.6049 [astro-ph.HE]}
  \BibitemShut {NoStop}%
\bibitem [{\citenamefont {Lopes}\ \emph {et~al.}(2023)\citenamefont {Lopes},
  \citenamefont {Marquez},\ and\ \citenamefont {Menezes}}]{Lopes:2022vjx}%
  \BibitemOpen
  \bibfield  {author} {\bibinfo {author} {\bibfnamefont {L.~L.}\ \bibnamefont
  {Lopes}}, \bibinfo {author} {\bibfnamefont {K.~D.}\ \bibnamefont {Marquez}},\
  and\ \bibinfo {author} {\bibfnamefont {D.~P.}\ \bibnamefont {Menezes}},\
  }\href {https://doi.org/10.1103/PhysRevD.107.036011} {\bibfield  {journal}
  {\bibinfo  {journal} {Phys. Rev. D}\ }\textbf {\bibinfo {volume} {107}},\
  \bibinfo {pages} {036011} (\bibinfo {year} {2023})},\ \Eprint
  {https://arxiv.org/abs/2211.17153} {arXiv:2211.17153 [hep-ph]} \BibitemShut
  {NoStop}%
\bibitem [{\citenamefont {{Sedrakian}}\ and\ \citenamefont
  {{Harutyunyan}}(2022)}]{sedrakian2022}%
  \BibitemOpen
  \bibfield  {author} {\bibinfo {author} {\bibfnamefont {A.}~\bibnamefont
  {{Sedrakian}}}\ and\ \bibinfo {author} {\bibfnamefont {A.}~\bibnamefont
  {{Harutyunyan}}},\ }\href {https://doi.org/10.1140/epja/s10050-022-00792-w}
  {\bibfield  {journal} {\bibinfo  {journal} {European Physical Journal A}\
  }\textbf {\bibinfo {volume} {58}},\ \bibinfo {eid} {137} (\bibinfo {year}
  {2022})},\ \Eprint {https://arxiv.org/abs/2202.12083} {arXiv:2202.12083
  [nucl-th]} \BibitemShut {NoStop}%
\bibitem [{\citenamefont {Li}\ and\ \citenamefont
  {Sedrakian}(2019)}]{Li:2019tjx}%
  \BibitemOpen
  \bibfield  {author} {\bibinfo {author} {\bibfnamefont {J.~J.}\ \bibnamefont
  {Li}}\ and\ \bibinfo {author} {\bibfnamefont {A.}~\bibnamefont {Sedrakian}},\
  }\href {https://doi.org/10.3847/2041-8213/ab1090} {\bibfield  {journal}
  {\bibinfo  {journal} {Astrophys. J. Lett.}\ }\textbf {\bibinfo {volume}
  {874}},\ \bibinfo {pages} {L22} (\bibinfo {year} {2019})},\ \Eprint
  {https://arxiv.org/abs/1904.02006} {arXiv:1904.02006 [nucl-th]} \BibitemShut
  {NoStop}%
\bibitem [{\citenamefont {Sch\"urhoff}\ \emph {et~al.}(2010)\citenamefont
  {Sch\"urhoff}, \citenamefont {Schramm},\ and\ \citenamefont
  {Dexheimer}}]{Schurhoff:2010ph}%
  \BibitemOpen
  \bibfield  {author} {\bibinfo {author} {\bibfnamefont {T.}~\bibnamefont
  {Sch\"urhoff}}, \bibinfo {author} {\bibfnamefont {S.}~\bibnamefont
  {Schramm}},\ and\ \bibinfo {author} {\bibfnamefont {V.}~\bibnamefont
  {Dexheimer}},\ }\href {https://doi.org/10.1088/2041-8205/724/1/L74}
  {\bibfield  {journal} {\bibinfo  {journal} {Astrophys. J. Lett.}\ }\textbf
  {\bibinfo {volume} {724}},\ \bibinfo {pages} {L74} (\bibinfo {year}
  {2010})},\ \Eprint {https://arxiv.org/abs/1008.0957} {arXiv:1008.0957
  [astro-ph.SR]} \BibitemShut {NoStop}%
\bibitem [{\citenamefont {Prakash}\ \emph {et~al.}(1992)\citenamefont
  {Prakash}, \citenamefont {Prakash}, \citenamefont {Lattimer},\ and\
  \citenamefont {Pethick}}]{Prakash:1992zng}%
  \BibitemOpen
  \bibfield  {author} {\bibinfo {author} {\bibfnamefont {M.}~\bibnamefont
  {Prakash}}, \bibinfo {author} {\bibfnamefont {M.}~\bibnamefont {Prakash}},
  \bibinfo {author} {\bibfnamefont {J.~M.}\ \bibnamefont {Lattimer}},\ and\
  \bibinfo {author} {\bibfnamefont {C.~J.}\ \bibnamefont {Pethick}},\ }\href
  {https://doi.org/10.1086/186376} {\bibfield  {journal} {\bibinfo  {journal}
  {Astrophys. J. Lett.}\ }\textbf {\bibinfo {volume} {390}},\ \bibinfo {pages}
  {L77} (\bibinfo {year} {1992})}\BibitemShut {NoStop}%
\bibitem [{\citenamefont {{Issifu}}\ \emph
  {et~al.}(2024{\natexlab{b}})\citenamefont {{Issifu}}, \citenamefont
  {{Thakur}}, \citenamefont {{da Silva}}, \citenamefont {{Marquez}},
  \citenamefont {{Menezes}}, \citenamefont {{Dutra}}, \citenamefont
  {{Louren{\c{c}}o}},\ and\ \citenamefont {{Frederico}}}]{Issifu:2024htq}%
  \BibitemOpen
  \bibfield  {author} {\bibinfo {author} {\bibfnamefont {A.}~\bibnamefont
  {{Issifu}}}, \bibinfo {author} {\bibfnamefont {P.}~\bibnamefont {{Thakur}}},
  \bibinfo {author} {\bibfnamefont {F.~M.}\ \bibnamefont {{da Silva}}},
  \bibinfo {author} {\bibfnamefont {K.~D.}\ \bibnamefont {{Marquez}}}, \bibinfo
  {author} {\bibfnamefont {D.~P.}\ \bibnamefont {{Menezes}}}, \bibinfo {author}
  {\bibfnamefont {M.}~\bibnamefont {{Dutra}}}, \bibinfo {author} {\bibfnamefont
  {O.}~\bibnamefont {{Louren{\c{c}}o}}},\ and\ \bibinfo {author} {\bibfnamefont
  {T.}~\bibnamefont {{Frederico}}},\ }\href
  {https://doi.org/10.48550/arXiv.2412.17946} {\bibfield  {journal} {\bibinfo
  {journal} {arXiv e-prints}\ ,\ \bibinfo {eid} {arXiv:2412.17946}} (\bibinfo
  {year} {2024}{\natexlab{b}})},\ \Eprint {https://arxiv.org/abs/2412.17946}
  {arXiv:2412.17946 [hep-ph]} \BibitemShut {NoStop}%
\bibitem [{\citenamefont {{Pfister}}(2007)}]{pfister2007GReGr}%
  \BibitemOpen
  \bibfield  {author} {\bibinfo {author} {\bibfnamefont {H.}~\bibnamefont
  {{Pfister}}},\ }\href {https://doi.org/10.1007/s10714-007-0521-4} {\bibfield
  {journal} {\bibinfo  {journal} {General Relativity and Gravitation}\ }\textbf
  {\bibinfo {volume} {39}},\ \bibinfo {pages} {1735} (\bibinfo {year}
  {2007})}\BibitemShut {NoStop}%
\bibitem [{\citenamefont {{Ciufolini}}\ and\ \citenamefont
  {{Pavlis}}(2004)}]{ciufolini2004Natur}%
  \BibitemOpen
  \bibfield  {author} {\bibinfo {author} {\bibfnamefont {I.}~\bibnamefont
  {{Ciufolini}}}\ and\ \bibinfo {author} {\bibfnamefont {E.~C.}\ \bibnamefont
  {{Pavlis}}},\ }\href {https://doi.org/10.1038/nature03007} {\bibfield
  {journal} {\bibinfo  {journal} {\nat}\ }\textbf {\bibinfo {volume} {431}},\
  \bibinfo {pages} {958} (\bibinfo {year} {2004})}\BibitemShut {NoStop}%
\bibitem [{\citenamefont {{Bonazzola}}\ \emph {et~al.}(1993)\citenamefont
  {{Bonazzola}}, \citenamefont {{Gourgoulhon}}, \citenamefont {{Salgado}},\
  and\ \citenamefont {{Marck}}}]{bonazzola1993A&A}%
  \BibitemOpen
  \bibfield  {author} {\bibinfo {author} {\bibfnamefont {S.}~\bibnamefont
  {{Bonazzola}}}, \bibinfo {author} {\bibfnamefont {E.}~\bibnamefont
  {{Gourgoulhon}}}, \bibinfo {author} {\bibfnamefont {M.}~\bibnamefont
  {{Salgado}}},\ and\ \bibinfo {author} {\bibfnamefont {J.~A.}\ \bibnamefont
  {{Marck}}},\ }\href@noop {} {\bibfield  {journal} {\bibinfo  {journal}
  {A\&A}\ }\textbf {\bibinfo {volume} {278}},\ \bibinfo {pages} {421} (\bibinfo
  {year} {1993})}\BibitemShut {NoStop}%
\bibitem [{\citenamefont {{Marques}}\ \emph {et~al.}(2017)\citenamefont
  {{Marques}}, \citenamefont {{Oertel}}, \citenamefont {{Hempel}},\ and\
  \citenamefont {{Novak}}}]{marques2017PhRvC}%
  \BibitemOpen
  \bibfield  {author} {\bibinfo {author} {\bibfnamefont {M.}~\bibnamefont
  {{Marques}}}, \bibinfo {author} {\bibfnamefont {M.}~\bibnamefont {{Oertel}}},
  \bibinfo {author} {\bibfnamefont {M.}~\bibnamefont {{Hempel}}},\ and\
  \bibinfo {author} {\bibfnamefont {J.}~\bibnamefont {{Novak}}},\ }\href
  {https://doi.org/10.1103/PhysRevC.96.045806} {\bibfield  {journal} {\bibinfo
  {journal} {\prc}\ }\textbf {\bibinfo {volume} {96}},\ \bibinfo {eid} {045806}
  (\bibinfo {year} {2017})},\ \Eprint {https://arxiv.org/abs/1706.02913}
  {arXiv:1706.02913 [nucl-th]} \BibitemShut {NoStop}%
\bibitem [{\citenamefont {{Komatsu}}\ \emph {et~al.}(1989)\citenamefont
  {{Komatsu}}, \citenamefont {{Eriguchi}},\ and\ \citenamefont
  {{Hachisu}}}]{komatsu1989MNRAS}%
  \BibitemOpen
  \bibfield  {author} {\bibinfo {author} {\bibfnamefont {H.}~\bibnamefont
  {{Komatsu}}}, \bibinfo {author} {\bibfnamefont {Y.}~\bibnamefont
  {{Eriguchi}}},\ and\ \bibinfo {author} {\bibfnamefont {I.}~\bibnamefont
  {{Hachisu}}},\ }\href {https://doi.org/10.1093/mnras/237.2.355} {\bibfield
  {journal} {\bibinfo  {journal} {\mnras}\ }\textbf {\bibinfo {volume} {237}},\
  \bibinfo {pages} {355} (\bibinfo {year} {1989})}\BibitemShut {NoStop}%
\bibitem [{\citenamefont {{Cook}}\ \emph {et~al.}(1994)\citenamefont {{Cook}},
  \citenamefont {{Shapiro}},\ and\ \citenamefont {{Teukolsky}}}]{cook1994ApJ}%
  \BibitemOpen
  \bibfield  {author} {\bibinfo {author} {\bibfnamefont {G.~B.}\ \bibnamefont
  {{Cook}}}, \bibinfo {author} {\bibfnamefont {S.~L.}\ \bibnamefont
  {{Shapiro}}},\ and\ \bibinfo {author} {\bibfnamefont {S.~A.}\ \bibnamefont
  {{Teukolsky}}},\ }\href {https://doi.org/10.1086/173721} {\bibfield
  {journal} {\bibinfo  {journal} {\apj}\ }\textbf {\bibinfo {volume} {422}},\
  \bibinfo {pages} {227} (\bibinfo {year} {1994})}\BibitemShut {NoStop}%
\bibitem [{\citenamefont {{Sorkin}}(1982)}]{1982ApJ...257..847S}%
  \BibitemOpen
  \bibfield  {author} {\bibinfo {author} {\bibfnamefont {R.~D.}\ \bibnamefont
  {{Sorkin}}},\ }\href {https://doi.org/10.1086/160034} {\bibfield  {journal}
  {\bibinfo  {journal} {\apj}\ }\textbf {\bibinfo {volume} {257}},\ \bibinfo
  {pages} {847} (\bibinfo {year} {1982})}\BibitemShut {NoStop}%
\bibitem [{\citenamefont {{Friedman}}\ \emph {et~al.}(1988)\citenamefont
  {{Friedman}}, \citenamefont {{Ipser}},\ and\ \citenamefont
  {{Sorkin}}}]{friedman1988ApJ}%
  \BibitemOpen
  \bibfield  {author} {\bibinfo {author} {\bibfnamefont {J.~L.}\ \bibnamefont
  {{Friedman}}}, \bibinfo {author} {\bibfnamefont {J.~R.}\ \bibnamefont
  {{Ipser}}},\ and\ \bibinfo {author} {\bibfnamefont {R.~D.}\ \bibnamefont
  {{Sorkin}}},\ }\href {https://doi.org/10.1086/166043} {\bibfield  {journal}
  {\bibinfo  {journal} {\apj}\ }\textbf {\bibinfo {volume} {325}},\ \bibinfo
  {pages} {722} (\bibinfo {year} {1988})}\BibitemShut {NoStop}%
\bibitem [{\citenamefont {{Kaplan}}\ \emph {et~al.}(2014)\citenamefont
  {{Kaplan}}, \citenamefont {{Ott}}, \citenamefont {{O'Connor}}, \citenamefont
  {{Kiuchi}}, \citenamefont {{Roberts}},\ and\ \citenamefont
  {{Duez}}}]{kaplan2014ApJ}%
  \BibitemOpen
  \bibfield  {author} {\bibinfo {author} {\bibfnamefont {J.~D.}\ \bibnamefont
  {{Kaplan}}}, \bibinfo {author} {\bibfnamefont {C.~D.}\ \bibnamefont {{Ott}}},
  \bibinfo {author} {\bibfnamefont {E.~P.}\ \bibnamefont {{O'Connor}}},
  \bibinfo {author} {\bibfnamefont {K.}~\bibnamefont {{Kiuchi}}}, \bibinfo
  {author} {\bibfnamefont {L.}~\bibnamefont {{Roberts}}},\ and\ \bibinfo
  {author} {\bibfnamefont {M.}~\bibnamefont {{Duez}}},\ }\href
  {https://doi.org/10.1088/0004-637X/790/1/19} {\bibfield  {journal} {\bibinfo
  {journal} {\apj}\ }\textbf {\bibinfo {volume} {790}},\ \bibinfo {eid} {19}
  (\bibinfo {year} {2014})},\ \Eprint {https://arxiv.org/abs/1306.4034}
  {arXiv:1306.4034 [astro-ph.HE]} \BibitemShut {NoStop}%
\bibitem [{\citenamefont {{Andersson}}\ and\ \citenamefont
  {{Kokkotas}}(2001)}]{andersson2001IJMPD}%
  \BibitemOpen
  \bibfield  {author} {\bibinfo {author} {\bibfnamefont {N.}~\bibnamefont
  {{Andersson}}}\ and\ \bibinfo {author} {\bibfnamefont {K.~D.}\ \bibnamefont
  {{Kokkotas}}},\ }\href {https://doi.org/10.1142/S0218271801001062} {\bibfield
   {journal} {\bibinfo  {journal} {International Journal of Modern Physics D}\
  }\textbf {\bibinfo {volume} {10}},\ \bibinfo {pages} {381} (\bibinfo {year}
  {2001})},\ \Eprint {https://arxiv.org/abs/gr-qc/0010102} {arXiv:gr-qc/0010102
  [gr-qc]} \BibitemShut {NoStop}%
\bibitem [{\citenamefont {{Cromartie}}\ \emph {et~al.}(2020)\citenamefont
  {{Cromartie}}, \citenamefont {{Fonseca}}, \citenamefont {{Ransom}} \emph
  {et~al.}}]{NANOGrav:2019jur}%
  \BibitemOpen
  \bibfield  {author} {\bibinfo {author} {\bibfnamefont {H.~T.}\ \bibnamefont
  {{Cromartie}}}, \bibinfo {author} {\bibfnamefont {E.}~\bibnamefont
  {{Fonseca}}}, \bibinfo {author} {\bibfnamefont {S.~M.}\ \bibnamefont
  {{Ransom}}}, \emph {et~al.},\ }\href
  {https://doi.org/10.1038/s41550-019-0880-2} {\bibfield  {journal} {\bibinfo
  {journal} {Nature Astronomy}\ }\textbf {\bibinfo {volume} {4}},\ \bibinfo
  {pages} {72} (\bibinfo {year} {2020})},\ \Eprint
  {https://arxiv.org/abs/1904.06759} {arXiv:1904.06759 [astro-ph.HE]}
  \BibitemShut {NoStop}%
\bibitem [{\citenamefont {{Hessels}}\ \emph {et~al.}(2006)\citenamefont
  {{Hessels}}, \citenamefont {{Ransom}}, \citenamefont {{Stairs}},
  \citenamefont {{Freire}}, \citenamefont {{Kaspi}},\ and\ \citenamefont
  {{Camilo}}}]{hessels2006Sci}%
  \BibitemOpen
  \bibfield  {author} {\bibinfo {author} {\bibfnamefont {J.~W.~T.}\
  \bibnamefont {{Hessels}}}, \bibinfo {author} {\bibfnamefont {S.~M.}\
  \bibnamefont {{Ransom}}}, \bibinfo {author} {\bibfnamefont {I.~H.}\
  \bibnamefont {{Stairs}}}, \bibinfo {author} {\bibfnamefont {P.~C.~C.}\
  \bibnamefont {{Freire}}}, \bibinfo {author} {\bibfnamefont {V.~M.}\
  \bibnamefont {{Kaspi}}},\ and\ \bibinfo {author} {\bibfnamefont
  {F.}~\bibnamefont {{Camilo}}},\ }\href
  {https://doi.org/10.1126/science.1123430} {\bibfield  {journal} {\bibinfo
  {journal} {Science}\ }\textbf {\bibinfo {volume} {311}},\ \bibinfo {pages}
  {1901} (\bibinfo {year} {2006})},\ \Eprint
  {https://arxiv.org/abs/astro-ph/0601337} {arXiv:astro-ph/0601337 [astro-ph]}
  \BibitemShut {NoStop}%
\bibitem [{\citenamefont {Janka}(2012)}]{Janka:2012wk}%
  \BibitemOpen
  \bibfield  {author} {\bibinfo {author} {\bibfnamefont {H.-T.}\ \bibnamefont
  {Janka}},\ }\href {https://doi.org/10.1146/annurev-nucl-102711-094901}
  {\bibfield  {journal} {\bibinfo  {journal} {Ann. Rev. Nucl. Part. Sci.}\
  }\textbf {\bibinfo {volume} {62}},\ \bibinfo {pages} {407} (\bibinfo {year}
  {2012})},\ \Eprint {https://arxiv.org/abs/1206.2503} {arXiv:1206.2503
  [astro-ph.SR]} \BibitemShut {NoStop}%
\bibitem [{\citenamefont {{Abbott}}\ \emph {et~al.}(2019)\citenamefont
  {{Abbott}}, \citenamefont {{Abbott}}, \citenamefont {{Abbott}}, \citenamefont
  {{Abraham}}, \citenamefont {{Acernese}} \emph {et~al.}}]{abbott2019PhRvX}%
  \BibitemOpen
  \bibfield  {author} {\bibinfo {author} {\bibfnamefont {B.~P.}\ \bibnamefont
  {{Abbott}}}, \bibinfo {author} {\bibfnamefont {R.}~\bibnamefont {{Abbott}}},
  \bibinfo {author} {\bibfnamefont {T.~D.}\ \bibnamefont {{Abbott}}}, \bibinfo
  {author} {\bibfnamefont {S.}~\bibnamefont {{Abraham}}}, \bibinfo {author}
  {\bibfnamefont {F.}~\bibnamefont {{Acernese}}}, \emph {et~al.},\ }\href
  {https://doi.org/10.1103/PhysRevX.9.031040} {\bibfield  {journal} {\bibinfo
  {journal} {Physical Review X}\ }\textbf {\bibinfo {volume} {9}},\ \bibinfo
  {eid} {031040} (\bibinfo {year} {2019})},\ \Eprint
  {https://arxiv.org/abs/1811.12907} {arXiv:1811.12907 [astro-ph.HE]}
  \BibitemShut {NoStop}%
\bibitem [{\citenamefont {{Bassa}}\ \emph {et~al.}(2017)\citenamefont
  {{Bassa}}, \citenamefont {{Pleunis}}, \citenamefont {{Hessels}},
  \citenamefont {{Ferrara}} \emph {et~al.}}]{bassa2017ApJ}%
  \BibitemOpen
  \bibfield  {author} {\bibinfo {author} {\bibfnamefont {C.~G.}\ \bibnamefont
  {{Bassa}}}, \bibinfo {author} {\bibfnamefont {Z.}~\bibnamefont {{Pleunis}}},
  \bibinfo {author} {\bibfnamefont {J.~W.~T.}\ \bibnamefont {{Hessels}}},
  \bibinfo {author} {\bibfnamefont {E.~C.}\ \bibnamefont {{Ferrara}}}, \emph
  {et~al.},\ }\href {https://doi.org/10.3847/2041-8213/aa8400} {\bibfield
  {journal} {\bibinfo  {journal} {\apjl}\ }\textbf {\bibinfo {volume} {846}},\
  \bibinfo {eid} {L20} (\bibinfo {year} {2017})},\ \Eprint
  {https://arxiv.org/abs/1709.01453} {arXiv:1709.01453 [astro-ph.HE]}
  \BibitemShut {NoStop}%
\bibitem [{\citenamefont {{Romani}}\ \emph {et~al.}(2022)\citenamefont
  {{Romani}}, \citenamefont {{Kandel}}, \citenamefont {{Filippenko}},
  \citenamefont {{Brink}},\ and\ \citenamefont {{Zheng}}}]{romani2022ApJ}%
  \BibitemOpen
  \bibfield  {author} {\bibinfo {author} {\bibfnamefont {R.~W.}\ \bibnamefont
  {{Romani}}}, \bibinfo {author} {\bibfnamefont {D.}~\bibnamefont {{Kandel}}},
  \bibinfo {author} {\bibfnamefont {A.~V.}\ \bibnamefont {{Filippenko}}},
  \bibinfo {author} {\bibfnamefont {T.~G.}\ \bibnamefont {{Brink}}},\ and\
  \bibinfo {author} {\bibfnamefont {W.}~\bibnamefont {{Zheng}}},\ }\href
  {https://doi.org/10.3847/2041-8213/ac8007} {\bibfield  {journal} {\bibinfo
  {journal} {\apjl}\ }\textbf {\bibinfo {volume} {934}},\ \bibinfo {eid} {L17}
  (\bibinfo {year} {2022})},\ \Eprint {https://arxiv.org/abs/2207.05124}
  {arXiv:2207.05124 [astro-ph.HE]} \BibitemShut {NoStop}%
\bibitem [{\citenamefont {{Khadkikar}}\ \emph {et~al.}(2021)\citenamefont
  {{Khadkikar}}, \citenamefont {{Raduta}}, \citenamefont {{Oertel}},\ and\
  \citenamefont {{Sedrakian}}}]{khadkikar2021PhRvC}%
  \BibitemOpen
  \bibfield  {author} {\bibinfo {author} {\bibfnamefont {S.}~\bibnamefont
  {{Khadkikar}}}, \bibinfo {author} {\bibfnamefont {A.~R.}\ \bibnamefont
  {{Raduta}}}, \bibinfo {author} {\bibfnamefont {M.}~\bibnamefont {{Oertel}}},\
  and\ \bibinfo {author} {\bibfnamefont {A.}~\bibnamefont {{Sedrakian}}},\
  }\href {https://doi.org/10.1103/PhysRevC.103.055811} {\bibfield  {journal}
  {\bibinfo  {journal} {\prc}\ }\textbf {\bibinfo {volume} {103}},\ \bibinfo
  {eid} {055811} (\bibinfo {year} {2021})},\ \Eprint
  {https://arxiv.org/abs/2102.00988} {arXiv:2102.00988 [astro-ph.HE]}
  \BibitemShut {NoStop}%
\bibitem [{\citenamefont {{Nunna}}\ \emph {et~al.}(2020)\citenamefont
  {{Nunna}}, \citenamefont {{Banik}},\ and\ \citenamefont
  {{Chatterjee}}}]{nunna2020ApJ}%
  \BibitemOpen
  \bibfield  {author} {\bibinfo {author} {\bibfnamefont {K.~P.}\ \bibnamefont
  {{Nunna}}}, \bibinfo {author} {\bibfnamefont {S.}~\bibnamefont {{Banik}}},\
  and\ \bibinfo {author} {\bibfnamefont {D.}~\bibnamefont {{Chatterjee}}},\
  }\href {https://doi.org/10.3847/1538-4357/ab8f2c} {\bibfield  {journal}
  {\bibinfo  {journal} {\apj}\ }\textbf {\bibinfo {volume} {896}},\ \bibinfo
  {eid} {109} (\bibinfo {year} {2020})},\ \Eprint
  {https://arxiv.org/abs/2002.07538} {arXiv:2002.07538 [astro-ph.HE]}
  \BibitemShut {NoStop}%
\bibitem [{\citenamefont {{Koliogiannis}}\ and\ \citenamefont
  {{Moustakidis}}(2021)}]{Koliogiannis:2020nhh}%
  \BibitemOpen
  \bibfield  {author} {\bibinfo {author} {\bibfnamefont {P.~S.}\ \bibnamefont
  {{Koliogiannis}}}\ and\ \bibinfo {author} {\bibfnamefont {C.~C.}\
  \bibnamefont {{Moustakidis}}},\ }\href
  {https://doi.org/10.3847/1538-4357/abe542} {\bibfield  {journal} {\bibinfo
  {journal} {\apj}\ }\textbf {\bibinfo {volume} {912}},\ \bibinfo {eid} {69}
  (\bibinfo {year} {2021})},\ \Eprint {https://arxiv.org/abs/2007.10424}
  {arXiv:2007.10424 [astro-ph.HE]} \BibitemShut {NoStop}%
\bibitem [{\citenamefont {{Lenka}}\ \emph {et~al.}(2019)\citenamefont
  {{Lenka}}, \citenamefont {{Char}},\ and\ \citenamefont
  {{Banik}}}]{Lenka:2018ehb}%
  \BibitemOpen
  \bibfield  {author} {\bibinfo {author} {\bibfnamefont {S.~S.}\ \bibnamefont
  {{Lenka}}}, \bibinfo {author} {\bibfnamefont {P.}~\bibnamefont {{Char}}},\
  and\ \bibinfo {author} {\bibfnamefont {S.}~\bibnamefont {{Banik}}},\ }\href
  {https://doi.org/10.1088/1361-6471/ab36a2} {\bibfield  {journal} {\bibinfo
  {journal} {Journal of Physics G Nuclear Physics}\ }\textbf {\bibinfo {volume}
  {46}},\ \bibinfo {eid} {105201} (\bibinfo {year} {2019})},\ \Eprint
  {https://arxiv.org/abs/1805.09492} {arXiv:1805.09492 [astro-ph.HE]}
  \BibitemShut {NoStop}%
\bibitem [{\citenamefont {{Thorne}}(1974)}]{thorne1974}%
  \BibitemOpen
  \bibfield  {author} {\bibinfo {author} {\bibfnamefont {K.~S.}\ \bibnamefont
  {{Thorne}}},\ }\href {https://doi.org/10.1086/152991} {\bibfield  {journal}
  {\bibinfo  {journal} {\apj}\ }\textbf {\bibinfo {volume} {191}},\ \bibinfo
  {pages} {507} (\bibinfo {year} {1974})}\BibitemShut {NoStop}%
\bibitem [{\citenamefont {{Morsink}}\ \emph {et~al.}(1999)\citenamefont
  {{Morsink}}, \citenamefont {{Stergioulas}},\ and\ \citenamefont
  {{Blattnig}}}]{morsink1999ApJ}%
  \BibitemOpen
  \bibfield  {author} {\bibinfo {author} {\bibfnamefont {S.~M.}\ \bibnamefont
  {{Morsink}}}, \bibinfo {author} {\bibfnamefont {N.}~\bibnamefont
  {{Stergioulas}}},\ and\ \bibinfo {author} {\bibfnamefont {S.~R.}\
  \bibnamefont {{Blattnig}}},\ }\href {https://doi.org/10.1086/306630}
  {\bibfield  {journal} {\bibinfo  {journal} {\apj}\ }\textbf {\bibinfo
  {volume} {510}},\ \bibinfo {pages} {854} (\bibinfo {year} {1999})},\ \Eprint
  {https://arxiv.org/abs/gr-qc/9806008} {arXiv:gr-qc/9806008 [gr-qc]}
  \BibitemShut {NoStop}%
\bibitem [{\citenamefont {{Koliogiannis}}\ and\ \citenamefont
  {{Moustakidis}}(2020)}]{koliogiannis2020PhRvC}%
  \BibitemOpen
  \bibfield  {author} {\bibinfo {author} {\bibfnamefont {P.~S.}\ \bibnamefont
  {{Koliogiannis}}}\ and\ \bibinfo {author} {\bibfnamefont {C.~C.}\
  \bibnamefont {{Moustakidis}}},\ }\href
  {https://doi.org/10.1103/PhysRevC.101.015805} {\bibfield  {journal} {\bibinfo
   {journal} {\prc}\ }\textbf {\bibinfo {volume} {101}},\ \bibinfo {eid}
  {015805} (\bibinfo {year} {2020})},\ \Eprint
  {https://arxiv.org/abs/1907.13375} {arXiv:1907.13375 [nucl-th]} \BibitemShut
  {NoStop}%
\bibitem [{\citenamefont {{Silva}}\ \emph {et~al.}(2021)\citenamefont
  {{Silva}}, \citenamefont {{Holgado}}, \citenamefont
  {{C{\'a}rdenas-Avenda{\~n}o}},\ and\ \citenamefont
  {{Yunes}}}]{silva2021PhRvL}%
  \BibitemOpen
  \bibfield  {author} {\bibinfo {author} {\bibfnamefont {H.~O.}\ \bibnamefont
  {{Silva}}}, \bibinfo {author} {\bibfnamefont {A.~M.}\ \bibnamefont
  {{Holgado}}}, \bibinfo {author} {\bibfnamefont {A.}~\bibnamefont
  {{C{\'a}rdenas-Avenda{\~n}o}}},\ and\ \bibinfo {author} {\bibfnamefont
  {N.}~\bibnamefont {{Yunes}}},\ }\href
  {https://doi.org/10.1103/PhysRevLett.126.181101} {\bibfield  {journal}
  {\bibinfo  {journal} {\prl}\ }\textbf {\bibinfo {volume} {126}},\ \bibinfo
  {eid} {181101} (\bibinfo {year} {2021})},\ \Eprint
  {https://arxiv.org/abs/2004.01253} {arXiv:2004.01253 [gr-qc]} \BibitemShut
  {NoStop}%
\bibitem [{\citenamefont {{Li}}\ \emph {et~al.}(2022)\citenamefont {{Li}},
  \citenamefont {{Wang}}, \citenamefont {{Wu}},\ and\ \citenamefont
  {{Wen}}}]{li2022CQGra}%
  \BibitemOpen
  \bibfield  {author} {\bibinfo {author} {\bibfnamefont {Y.}~\bibnamefont
  {{Li}}}, \bibinfo {author} {\bibfnamefont {J.}~\bibnamefont {{Wang}}},
  \bibinfo {author} {\bibfnamefont {Z.}~\bibnamefont {{Wu}}},\ and\ \bibinfo
  {author} {\bibfnamefont {D.}~\bibnamefont {{Wen}}},\ }\href
  {https://doi.org/10.1088/1361-6382/ac45d9} {\bibfield  {journal} {\bibinfo
  {journal} {Classical and Quantum Gravity}\ }\textbf {\bibinfo {volume}
  {39}},\ \bibinfo {eid} {035014} (\bibinfo {year} {2022})}\BibitemShut
  {NoStop}%
\bibitem [{\citenamefont {{Bejger}}\ \emph {et~al.}(2005)\citenamefont
  {{Bejger}}, \citenamefont {{Bulik}},\ and\ \citenamefont
  {{Haensel}}}]{bejger2005MNRAS}%
  \BibitemOpen
  \bibfield  {author} {\bibinfo {author} {\bibfnamefont {M.}~\bibnamefont
  {{Bejger}}}, \bibinfo {author} {\bibfnamefont {T.}~\bibnamefont {{Bulik}}},\
  and\ \bibinfo {author} {\bibfnamefont {P.}~\bibnamefont {{Haensel}}},\ }\href
  {https://doi.org/10.1111/j.1365-2966.2005.09575.x} {\bibfield  {journal}
  {\bibinfo  {journal} {\mnras}\ }\textbf {\bibinfo {volume} {364}},\ \bibinfo
  {pages} {635} (\bibinfo {year} {2005})},\ \Eprint
  {https://arxiv.org/abs/astro-ph/0508105} {arXiv:astro-ph/0508105 [astro-ph]}
  \BibitemShut {NoStop}%
\bibitem [{\citenamefont {{Burgay}}\ \emph {et~al.}(2003)\citenamefont
  {{Burgay}}, \citenamefont {{D'Amico}}, \citenamefont {{Possenti}},
  \citenamefont {{Manchester}} \emph {et~al.}}]{burgay2003Natur}%
  \BibitemOpen
  \bibfield  {author} {\bibinfo {author} {\bibfnamefont {M.}~\bibnamefont
  {{Burgay}}}, \bibinfo {author} {\bibfnamefont {N.}~\bibnamefont {{D'Amico}}},
  \bibinfo {author} {\bibfnamefont {A.}~\bibnamefont {{Possenti}}}, \bibinfo
  {author} {\bibfnamefont {R.~N.}\ \bibnamefont {{Manchester}}}, \emph
  {et~al.},\ }\href {https://doi.org/10.1038/nature02124} {\bibfield  {journal}
  {\bibinfo  {journal} {\nat}\ }\textbf {\bibinfo {volume} {426}},\ \bibinfo
  {pages} {531} (\bibinfo {year} {2003})},\ \Eprint
  {https://arxiv.org/abs/astro-ph/0312071} {arXiv:astro-ph/0312071 [astro-ph]}
  \BibitemShut {NoStop}%
\bibitem [{\citenamefont {{Kramer}}\ and\ \citenamefont
  {{Wex}}(2009)}]{kramer2009CQGra}%
  \BibitemOpen
  \bibfield  {author} {\bibinfo {author} {\bibfnamefont {M.}~\bibnamefont
  {{Kramer}}}\ and\ \bibinfo {author} {\bibfnamefont {N.}~\bibnamefont
  {{Wex}}},\ }\href {https://doi.org/10.1088/0264-9381/26/7/073001} {\bibfield
  {journal} {\bibinfo  {journal} {Classical and Quantum Gravity}\ }\textbf
  {\bibinfo {volume} {26}},\ \bibinfo {eid} {073001} (\bibinfo {year}
  {2009})}\BibitemShut {NoStop}%
\bibitem [{\citenamefont {{Morrison}}\ \emph {et~al.}(2004)\citenamefont
  {{Morrison}}, \citenamefont {{Baumgarte}}, \citenamefont {{Shapiro}},\ and\
  \citenamefont {{Pandharipande}}}]{morrison2004ApJ}%
  \BibitemOpen
  \bibfield  {author} {\bibinfo {author} {\bibfnamefont {I.~A.}\ \bibnamefont
  {{Morrison}}}, \bibinfo {author} {\bibfnamefont {T.~W.}\ \bibnamefont
  {{Baumgarte}}}, \bibinfo {author} {\bibfnamefont {S.~L.}\ \bibnamefont
  {{Shapiro}}},\ and\ \bibinfo {author} {\bibfnamefont {V.~R.}\ \bibnamefont
  {{Pandharipande}}},\ }\href {https://doi.org/10.1086/427235} {\bibfield
  {journal} {\bibinfo  {journal} {\apjl}\ }\textbf {\bibinfo {volume} {617}},\
  \bibinfo {pages} {L135} (\bibinfo {year} {2004})},\ \Eprint
  {https://arxiv.org/abs/astro-ph/0411353} {arXiv:astro-ph/0411353 [astro-ph]}
  \BibitemShut {NoStop}%
\bibitem [{\citenamefont {{Lattimer}}\ and\ \citenamefont
  {{Schutz}}(2005)}]{lattimer2005ApJ}%
  \BibitemOpen
  \bibfield  {author} {\bibinfo {author} {\bibfnamefont {J.~M.}\ \bibnamefont
  {{Lattimer}}}\ and\ \bibinfo {author} {\bibfnamefont {B.~F.}\ \bibnamefont
  {{Schutz}}},\ }\href {https://doi.org/10.1086/431543} {\bibfield  {journal}
  {\bibinfo  {journal} {\apj}\ }\textbf {\bibinfo {volume} {629}},\ \bibinfo
  {pages} {979} (\bibinfo {year} {2005})},\ \Eprint
  {https://arxiv.org/abs/astro-ph/0411470} {arXiv:astro-ph/0411470 [astro-ph]}
  \BibitemShut {NoStop}%
\bibitem [{\citenamefont {{Landry}}\ and\ \citenamefont
  {{Kumar}}(2018)}]{landry2018ApJ}%
  \BibitemOpen
  \bibfield  {author} {\bibinfo {author} {\bibfnamefont {P.}~\bibnamefont
  {{Landry}}}\ and\ \bibinfo {author} {\bibfnamefont {B.}~\bibnamefont
  {{Kumar}}},\ }\href {https://doi.org/10.3847/2041-8213/aaee76} {\bibfield
  {journal} {\bibinfo  {journal} {\apjl}\ }\textbf {\bibinfo {volume} {868}},\
  \bibinfo {eid} {L22} (\bibinfo {year} {2018})},\ \Eprint
  {https://arxiv.org/abs/1807.04727} {arXiv:1807.04727 [gr-qc]} \BibitemShut
  {NoStop}%
\bibitem [{\citenamefont {{Lim}}\ \emph {et~al.}(2019)\citenamefont {{Lim}},
  \citenamefont {{Holt}},\ and\ \citenamefont {{Stahulak}}}]{lim2019PhRvC}%
  \BibitemOpen
  \bibfield  {author} {\bibinfo {author} {\bibfnamefont {Y.}~\bibnamefont
  {{Lim}}}, \bibinfo {author} {\bibfnamefont {J.~W.}\ \bibnamefont {{Holt}}},\
  and\ \bibinfo {author} {\bibfnamefont {R.~J.}\ \bibnamefont {{Stahulak}}},\
  }\href {https://doi.org/10.1103/PhysRevC.100.035802} {\bibfield  {journal}
  {\bibinfo  {journal} {\prc}\ }\textbf {\bibinfo {volume} {100}},\ \bibinfo
  {eid} {035802} (\bibinfo {year} {2019})},\ \Eprint
  {https://arxiv.org/abs/1810.10992} {arXiv:1810.10992 [nucl-th]} \BibitemShut
  {NoStop}%
\bibitem [{\citenamefont {{Miao}}\ \emph {et~al.}(2022)\citenamefont {{Miao}},
  \citenamefont {{Li}},\ and\ \citenamefont {{Dai}}}]{miao2022MNRAS}%
  \BibitemOpen
  \bibfield  {author} {\bibinfo {author} {\bibfnamefont {Z.}~\bibnamefont
  {{Miao}}}, \bibinfo {author} {\bibfnamefont {A.}~\bibnamefont {{Li}}},\ and\
  \bibinfo {author} {\bibfnamefont {Z.-G.}\ \bibnamefont {{Dai}}},\ }\href
  {https://doi.org/10.1093/mnras/stac2015} {\bibfield  {journal} {\bibinfo
  {journal} {\mnras}\ }\textbf {\bibinfo {volume} {515}},\ \bibinfo {pages}
  {5071} (\bibinfo {year} {2022})},\ \Eprint {https://arxiv.org/abs/2107.07979}
  {arXiv:2107.07979 [astro-ph.HE]} \BibitemShut {NoStop}%
\bibitem [{\citenamefont {{Steiner}}\ \emph {et~al.}(2015)\citenamefont
  {{Steiner}}, \citenamefont {{Gandolfi}}, \citenamefont {{Fattoyev}},\ and\
  \citenamefont {{Newton}}}]{steiner2015}%
  \BibitemOpen
  \bibfield  {author} {\bibinfo {author} {\bibfnamefont {A.~W.}\ \bibnamefont
  {{Steiner}}}, \bibinfo {author} {\bibfnamefont {S.}~\bibnamefont
  {{Gandolfi}}}, \bibinfo {author} {\bibfnamefont {F.~J.}\ \bibnamefont
  {{Fattoyev}}},\ and\ \bibinfo {author} {\bibfnamefont {W.~G.}\ \bibnamefont
  {{Newton}}},\ }\href {https://doi.org/10.1103/PhysRevC.91.015804} {\bibfield
  {journal} {\bibinfo  {journal} {\prc}\ }\textbf {\bibinfo {volume} {91}},\
  \bibinfo {eid} {015804} (\bibinfo {year} {2015})},\ \Eprint
  {https://arxiv.org/abs/1403.7546} {arXiv:1403.7546 [nucl-th]} \BibitemShut
  {NoStop}%
\bibitem [{\citenamefont {{Breu}}\ and\ \citenamefont
  {{Rezzolla}}(2016)}]{breu2016}%
  \BibitemOpen
  \bibfield  {author} {\bibinfo {author} {\bibfnamefont {C.}~\bibnamefont
  {{Breu}}}\ and\ \bibinfo {author} {\bibfnamefont {L.}~\bibnamefont
  {{Rezzolla}}},\ }\href {https://doi.org/10.1093/mnras/stw575} {\bibfield
  {journal} {\bibinfo  {journal} {\mnras}\ }\textbf {\bibinfo {volume} {459}},\
  \bibinfo {pages} {646} (\bibinfo {year} {2016})},\ \Eprint
  {https://arxiv.org/abs/1601.06083} {arXiv:1601.06083 [gr-qc]} \BibitemShut
  {NoStop}%
\bibitem [{\citenamefont {{Ravenhall}}\ and\ \citenamefont
  {{Pethick}}(1994)}]{ravenhall1994ApJ}%
  \BibitemOpen
  \bibfield  {author} {\bibinfo {author} {\bibfnamefont {D.~G.}\ \bibnamefont
  {{Ravenhall}}}\ and\ \bibinfo {author} {\bibfnamefont {C.~J.}\ \bibnamefont
  {{Pethick}}},\ }\href {https://doi.org/10.1086/173935} {\bibfield  {journal}
  {\bibinfo  {journal} {\apj}\ }\textbf {\bibinfo {volume} {424}},\ \bibinfo
  {pages} {846} (\bibinfo {year} {1994})}\BibitemShut {NoStop}%
\bibitem [{\citenamefont {{Lattimer}}\ and\ \citenamefont
  {{Prakash}}(2001)}]{lattimer2001ApJ}%
  \BibitemOpen
  \bibfield  {author} {\bibinfo {author} {\bibfnamefont {J.~M.}\ \bibnamefont
  {{Lattimer}}}\ and\ \bibinfo {author} {\bibfnamefont {M.}~\bibnamefont
  {{Prakash}}},\ }\href {https://doi.org/10.1086/319702} {\bibfield  {journal}
  {\bibinfo  {journal} {\apj}\ }\textbf {\bibinfo {volume} {550}},\ \bibinfo
  {pages} {426} (\bibinfo {year} {2001})},\ \Eprint
  {https://arxiv.org/abs/astro-ph/0002232} {arXiv:astro-ph/0002232 [astro-ph]}
  \BibitemShut {NoStop}%
\bibitem [{\citenamefont {{Greif}}\ \emph {et~al.}(2020)\citenamefont
  {{Greif}}, \citenamefont {{Hebeler}}, \citenamefont {{Lattimer}},
  \citenamefont {{Pethick}},\ and\ \citenamefont {{Schwenk}}}]{greif2020ApJ}%
  \BibitemOpen
  \bibfield  {author} {\bibinfo {author} {\bibfnamefont {S.~K.}\ \bibnamefont
  {{Greif}}}, \bibinfo {author} {\bibfnamefont {K.}~\bibnamefont {{Hebeler}}},
  \bibinfo {author} {\bibfnamefont {J.~M.}\ \bibnamefont {{Lattimer}}},
  \bibinfo {author} {\bibfnamefont {C.~J.}\ \bibnamefont {{Pethick}}},\ and\
  \bibinfo {author} {\bibfnamefont {A.}~\bibnamefont {{Schwenk}}},\ }\href
  {https://doi.org/10.3847/1538-4357/abaf55} {\bibfield  {journal} {\bibinfo
  {journal} {\apj}\ }\textbf {\bibinfo {volume} {901}},\ \bibinfo {eid} {155}
  (\bibinfo {year} {2020})},\ \Eprint {https://arxiv.org/abs/2005.14164}
  {arXiv:2005.14164 [astro-ph.HE]} \BibitemShut {NoStop}%
\end{thebibliography}%

\end{document}